\documentclass[fleqn,12pt]{article}
\usepackage{gc,amssymb,cite}
\usepackage{graphics,epsf}
\usepackage[dvips]{graphicx}

\def\kappa{\varkappa}

\def\mn{_{\mu\nu}}
\def\MN{^{\mu\nu}}

\def\M{{\mathbb M}}
\def\R{{\mathbb R}}
\def\S{{\mathbb S}}

\def\eff{_{\rm eff}}

\def\AdS{anti-de Sitter}
\def\RN{Reissner-Nordstr\"om}

\def\sph{spherically symmetric}

\def\bh{black hole}

\def\bw{brane world}

\heads{K.A. Bronnikov and B.E. Meierovich}
       {Global strings in extra dimensions: a full map of solutions,
        trapping and the hierarchy problem}

\begin{document}
\onecol
\prepno{Arxiv: 0708.3439}{}

\Title
 {Global strings in extra dimensions: \yy a full map of solutions,\yy
    matter trapping, and the hierarchy problem}

\Authors{K.A. Bronnikov\foom 1} {and B.E. Meierovich\foom 2}
        {\small Center for Gravitation and Fundamental
             Metrology, VNIIMS,\\ 46 Ozyornaya St., Moscow 119361, Russia;\\
    Institute of Gravitation and Cosmology, PFUR,
             6 Miklukho-Maklaya St., Moscow 117198, Russia}
    {\small Kapitza Institute for Physical Problems,
         2 Kosygina St., Moscow 117334, Russia}

\Abstract
    {We consider ($d_0+2$)-dimensional configurations with global strings in
    the two extra dimensions and a flat metric in $d_0$ dimensions, endowed
    with a warp factor $e^{2\gamma}$ depending on the distance $l$ from the
    string center. All possible regular solutions to the field equations are
    classified by the behavior of the warp factor and the extra-dimensional
    circular radius $r(l)$. Solutions with $r\to \infty$ and $r\to \const
    >0$ as $l\to \infty$ are interpreted in terms of thick brane world
    models. Solutions with $r \to 0$ as $l \to l_c >0$, i.e., those with a
    second center, are interpreted as either multi-brane systems (which is
    appropriate for large enough distances $l_c$ between the centers) or as
    Kaluza-Klein type configurations with extra dimensions invisible due to
    their smallness. For the case of the Mexican-hat symmetry breaking
    potential, we build a full map of regular solutions on the
    ($\eps,\,\Gamma$) parameter plane, where $\eps$ acts as an effective
    cosmological constant while $\Gamma$ characterizes the gravitational
    field strength. The trapping properties of candidate brane worlds for
    test scalar fields are discussed. Good trapping properties for massive
    fields are found for models with growing warp factors. Kaluza-Klein type
    models are shown to possess nontrivial warp factor behaviors, leading to
    matter particle mass spectra which seem promising from the standpoint of
    hierarchy problems.  }



\section{Introduction}

  The multidimensional gravity concept, tracing back to the pioneering
  papers by Kaluza and Klein \cite{KK}, initially assumed that the extra
  dimensions remain unobservable due to their extreme smallness. Another
  class of multidimensional theories has been put forward in the 80s, on the
  basis of the idea that we live on a distinguished surface (brane) embedded
  in a higher-dimensional manifold, called the bulk \cite{brane}. This
  idea has recently become very popular in attempts to find an approach to
  a number of fundamental physical problems. The brane-world concept is
  broadly discussed in connection with the recent developments in
  supersymmetric string/M-theories \cite{M-theory}. The simple RS1 model
  \cite{RS1} has continued numerous attempts \cite{hierarchy} to find the
  origin of the enormous hierarchy of energy/mass scales observed in nature,
  which is a long-standing problem in particle physics. In astrophysics and
  cosmology, there are attempts to explain the dark matter and dark energy
  effects, to describe black holes and possible variation of fundamental
  constants, the CMB anisotropy etc.

  A great variety of brane world models may be found in the literature:
  branes in 5 or more dimensions, single or multiple branes, flat or curved
  branes, flat or curved bulk, compact or non-compact bulk (i.e., large or
  infinite extra dimensions), thin or thick branes, various symmetries of
  both bulks and branes, different kinds of matter forming the brane etc.

  In our view, the most natural physical idea leading to the emergence of
  distinguished surfaces in the space-time manifold is the idea of a phase
  transition with spontaneous symmetry breaking (SSB) which has already led
  to great success in many areas of physics and cosmology. In other words,
  it is reasonable to consider the brane world as a result of a phase
  transition at a very early stage of the Universe evolution. The existing
  macroscopic theory of phase transitions with SSB allows one to consider
  the brane world concept self-consistently and maximally escape the
  influence of model assumptions even without a detailed knowledge of the
  nature of physical vacuum. A necessary consequence of such phase
  transitions is the appearance of topological defects.

  Let us recall that a decisive step toward cosmological applications of the
  SSB concept was made in 1972 by Kirzhnits \cite{Kirzhnitz}. He assumed
  that, as in the case of solid substances, a symmetry of a field system,
  existing at sufficiently high temperatures, could be spontaneously broken
  as the temperature falls down. The first quantitative analysis of the
  cosmological consequences of SSB was given by Zel'dovich, Kobzarev, and
  Okun' \cite{Zel'dovich}.

  The properties of global topological defects are generally described with
  the aid of a multiplet of scalar fields playing the role of an order
  parameter. If a defect is to be interpreted as the origin of a brane world,
  its structure is determined by the self-gravity of a scalar field system
  and may be described by a set of Einstein and scalar equations.

  This approach to the brane-world concept has been employed by many authors
  for construction of thick branes in five (see, e.g., \cite{wolfe, thick-5,
  melfo, ghoroku, soda, thick2, thick3}) and more (see, e.g.,
  \cite{vil, shap, we-05, we-06} and references therein)  dimensions.

  In our previous papers \cite{thick2, thick3, we-05, we-06}, we have
  analyzed the gravitational properties of candidate (thick) brane worlds
  with the $d_0$-dimensional Minkowski metric and global topological defects
  in $d_1+1$ extra dimensions. Our general formulation covered such
  particular cases as a brane (domain wall) in five-dimensional spacetime
  (one extra dimension), a global cosmic string with winding number $n = 1$
  (two extra dimensions), and global monopoles (three or more extra
  dimensions). We restricted ourselves to Minkowski branes since most of the
  existing problems are clearly seen even in these comparatively simple
  systems; on the other hand, in the majority of physical situations, the
  intrinsic curvature of the brane itself is much smaller than the curvature
  related to brane formation, and therefore the main qualitative features of
  Minkowski branes should survive in realistic curved branes.

  Our treatment differed from many others (e.g., \cite{vil, shap}) in that
  we have considered all kinds of regular solutions to the corresponding
  field equations, including those with growing warp factors, and have
  emphasized good trapping properties of the latter.

  We have shown, in particular, that there are as many as seven classes of
  regular solutions of the field equations describing global strings and
  monopoles in extra dimensions; two of them exist for monopoles only while
  the other five are found for both strings and monopoles. Some of these
  configurations have exponentially growing warp factors ($\e^{2\gamma}$ in
  the metric (\ref{ds}), see below) at large distances from the core. They
  are shown to trap linear test scalar fields of any mass and momentum.
  Others, ending with a flat metric, have a warp factor tending to a
  constant value, determined by the shape of the symmetry breaking
  potential. They are also shown to trap test scalar fields with masses
  bounded above by a value depending on the particular parameters of
  the topological defect.

  Though the general classification of \cite{we-05, we-06} covers all
  possible regular configuration, the important question of location of
  different solutions in the space of physical parameters remained open. One
  of the goals of this paper is to answer this question for the particular
  case of a global string in two extra dimensions and a Mexican-hat symmetry
  breaking potential. The problem then contains two essential physical
  parameters: $\eps$, the dimensionless cosmological constant, and $\Gamma$,
  characterizing the gravitational field strength. The border lines in the
  ($\eps,\,\Gamma$) plane, separating different classes of regular solutions
  (those extending to infinite circular radii $r$, those with a cylindrical
  geometry far from the center and those with with two regular centers), are
  found numerically, and the asymptotic dependences $\eps(\Gamma)$ as
  $\Gamma \to 0$ and $\Gamma \to \infty$ are derived analytically.

  Another goal is to give a more complete description of the configurations
  of interest described by these solutions. Thus, we describe the trapping
  properties of different thick brane world models for classical particles,
  scalar fields and gravity. We also argue that one of the classes of
  regular configurations, those with two centers, can lead to promising
  models with nontrivial particle mass spectra. The point is that the warp
  factors of such configurations can have several minima at different
  levels, where test particles and fields may be gravitationally trapped
  with different energies. Though, in this case, one should abandon the
  brane world concept and admit that the extra dimensions are invisible due
  to their smallness, i.e., interpret the solutions in the spirit of
  Kaluza-Klein theories.

  There is a growing number of publications devoted to brane worlds with two
  extra dimensions, see \cite{d1=1} and references therein. In many cases
  the results are obtained numerically using simplified models with
  specially chosen sets of parameters. In our macroscopic approach, based
  on the theory of phase transitions with SSB, we try to reduce the
  influence of model assumptions to a minimum and to cover the full range
  of opportunities. Our main result, the full map of regular solutions for
  a system with the Mexican-hat symmetry-breaking potential $V(\phi)$,
  should probably retain its basic qualitative features for other potentials
  with a similar disposition of extremum points.

  The paper is organized as follows. \sect 2 outlines the problem setting,
  including the geometry, field equations and boundary conditions providing
  space-time regularity. \sect 3 describes the simplest solutions with a
  constant scalar field, needed for comparison in what follows. \sect 4
  is central in the paper: we give a general description and classification
  of possible regular solutions on the basis of our previous work
  \cite{we-05, we-06} and present a map showing the location of different
  solutions with the Mexican hat potential in the parameter space of the
  problem. In Sections 5 and 6 we describe some further details of the
  solutions and give an analytic derivation of the asymptotic behavior of
  the curves drawn in the map. \sect 8 discusses the trapping properties of
  thick branes described by the above solutions. \sect 9 outlines the
  properties of configurations with two centers, and \sect 10 is a
  conclusion.

\section{Problem setting}

\subsection{Geometry and regularity conditions}

  Our main interest here is 6D space-time with a cosmic string in the two
  extra dimensions. Let us, however, begin with a more general geometry:
  consider a ($D = d_0 + d_1 +1)$-dimensional space-time with the
  structure $\M^{d_0 }\times \R_u \times \S^{d_1 }$ and the metric
\beq
     ds^2 = \e^{2\gamma (u)}\eta _{\mu \nu }dx^{\mu }dx^{\nu }
       -\left( \e^{2\alpha(u)}du^2 + \e^{2\beta (u)}d\Omega^2 \right) .
                                           \label{ds}
\eeq
  where $\eta _{\mu \nu }  = \diag (1,-1,...,-1)$ is the $d_0$-dimensional
  Minkowski metric ($d_0 > 1$), $d\Omega$ is a linear element on a
  $d_1$-dimensional unit sphere $\S^{d_1}$, $\alpha $, $\beta $ and $\gamma
  $ are functions of the radial coordinate $u$ with the definition domain
  $\R_u \subseteq \R$ to be specified later. We will also use the notation
  $r \equiv \e^\beta$ where $r$ is the spherical (circular for $d_1=1$)
  radius.

  The Riemann tensor $R^{AB}{}_{CD}$ is diagonal with respect to pairs of
  indices and has the nonzero components
\bear                                   \label{Riem}
    R_{\ \ \rho \sigma }^{\mu \nu }
     \eql -\e^{-2\alpha }\gamma'{}^2\delta\MN_{\ \ \rho \sigma },
\nn
    R_{\ \ cd}^{ab} \eql (\e^{-2\beta }-\e^{-2\alpha }\beta'{}^2)
            \delta _{\ \ cd}^{ab},
\nn
    R_{\ \ u\nu }^{u\mu } \eql -\delta _{\nu }^{\mu}
            \e^{-\gamma -\alpha}( \e^{\gamma -\alpha }\gamma')',
\nn
    R_{\ \ ub}^{ua}  \eql -\delta_{b}^{a}
            \e^{-\beta -\alpha }( \e^{\beta -\alpha }\beta')',
\nn
    R_{\ \ b\nu }^{a\mu } \eql -\delta _{\nu}^{\mu }
        \delta _{b}^{a}\e^{-2\alpha}\gamma' \beta'.
\ear
  where $\delta _{\ \ \rho \sigma }^{\mu \nu} =\delta _{\rho }^{\mu }
  \delta _{\sigma }^{\nu }-\delta _{\sigma}^{\mu }\delta _{\rho }^{\nu }$
  and similarly for other indices.  The indices $\mu,\nu ,...$ correspond
  to $d_0 $-dimensional (physical) space-time, $a,b,...$ to $d_1 $ angular
  coordinates on $\S^{d_1 }$, and $A,B,...$ to all $D$ coordinates.

  A necessary condition of regularity is finiteness of all algebraic
  invariants of the Riemann tensor. In our case it is sufficient to deal
  with  the Kretschmann scalar $K=R_{\ \ CD}^{AB}R_{\ \ AB}^{CD}$ since it
  is a sum of squares of all nonzero $R_{{\rm  \ \ }CD}^{AB}$. Hence, in
  regular configurations all components of the Riemann tensor (\ref{Riem})
  are finite.

  In the Gaussian gauge $\alpha =0$, $u = l$ being the proper distance
  along the radial direction, the regularity conditions at $r>0$ look very
  simple:
\beq
    \beta' ,\quad \beta'' ,\quad \gamma ',\quad \gamma''
    \qquad \mbox{should be finite.}             \label{Reg-Gauss}
\eeq
  The regularity conditions at the center $r=0$ follow from finiteness of
  the Riemann tensor components $R_{\ \ cd}^{ab}$ and coincide with the
  regular center conditions in usual static, spherically symmetric metrics.
  In terms of an arbitrary $u$ coordinate, the regular center conditions are
\beq
    \gamma =\gamma _{c} + O(r^2) \qquad
    \e^{\beta -\alpha }|\beta '| = 1 +O(r^2) \qquad {\rm as \ \ }r\to 0.
                                \label{Reg center}
\eeq
  The latter condition means a correct ($= 2\pi$) circumference to radius
  ratio, or, equivalently, $dr^2 = dl^2$; $\gamma_c$ is a constant which can
  be set to zero by rescaling the coordinates $x^\mu$.

  The string case $d_1 =1$ has a specific feature: there is only one
  angular coordinate, therefore $\delta _{\ \ cd}^{ab}\equiv 0$, hence
  $R^{ab}{}_{cd} \equiv 0$. However, a conical singularity is possible
  (i.e., an angular deficit, $dr^2 <dl^2$, or excess, $dr^2 > dl^2$), which
  is a pointwise, delta-like curvature peak over this zero level, as in the
  case of an ordinary cone top. Its existence actually means that there is
  some pointlike object with respect to the two extra dimensions, or a thin
  brane in the space-time as a whole.

  Below we will consider entirely regular configurations, excluding,
  among others, conical singularities.

\subsection{Topological defects. Field equations}

   A global defect with a nonzero topological charge can be constructed with
   a multiplet of $d_1+1$ real scalar fields $\phi^k$, in the same way as,
   e.g., in \cite {we-05}. It comprises a ``hedgehog'' configuration in $\R_u
   \times \S^{d_1}$:
\[
    \phi^k = \phi (u) n^k (x^a),
\]
   $n^k$ is a unit vector in the $d_1+1$-dimensional Euclidean target space
   of the scalar fields: $n^k n^k =1$.

   The total Lagrangian of the system is taken in the form
\beq                                                         \label{Lagr}
    L = \frac{R}{2\kappa^2}
                + \Half g^{AB}\d_A\phi^k \d_B \phi^k - V (\phi),
\eeq
   where $R$ is the $D$-dimensional scalar curvature, $\kappa^2 $ is the
   $D$-dimensional gravitational constant, and $V$ is a symmetry-breaking
   potential depending on $\phi^2 (u) = \phi^{a} \phi^{a}$.

   The case $d_1 = 0$, with only one extra dimension, is a flat domain wall.
   Regular thick Minkowski branes supported by scalar fields with arbitrary
   potentials were analyzed in \cite{thick2, thick3}.

   The case $d_1 = 1$ is a global cosmic string with the winding number
   $n=1$, to be discussed here in detail. In case $d_1 > 2$ we have a global
   monopole in the extra space-like dimensions, see, e.g., \cite{vil, shap,
   Benson-Cho, Cho-Vil, we-05, we-06}.

   Let us write down the scalar field equation and three components of the
   Einstein equations for such systems in the Gaussian gauge $\alpha =0$, $u
   = l$ (the prime denotes $d/dl$):
\bear
    \phi'' +( d_0 \gamma ' +d_1 \beta'{}^2) \phi '
    -d_1 \e^{-2\beta }\phi \eql \frac{\d V}{\d \phi},    \label{phi''}
\\
     \gamma'' +d_0 \gamma '{}^2 +d_1 \beta' \gamma'
            \eql -\frac{2\kappa^2}{D-2}V,       \label{gamma''}
\\
    \beta'' +d_0 \beta' \gamma' +d_1 \beta '{}^2
    \eql (d_1 -1-k^2 \phi^2 )\e^{-2\beta }-\frac{2\kappa^2 }{D-2}V,
                                        \label{beta''}
\\
    ( d_1 \beta' +d_0 \gamma')^2 -d_0 \gamma'{}^2
        -d_1 \beta '{}^2 \eql \kappa^2
    (\phi'{}^2-2V) +d_1 \e^{-2\beta}(d_1 -1-\kappa^2 \phi^2).
                                \label{int}
\ear
   Any three of the above four equations are independent, and the fourth one
   is their consequence.

\subsection{Boundary conditions and fine-tuning relations}

   The metric can be rewritten in the form
\beq
     ds^2 =\e^{2\gamma (l)}\eta\mn dx^{\mu }dx^{\nu}-dl^2
                -r^2 (l) d\Omega^2,         \label{Metric}
\eeq
   where $r = \e^{\beta }$ is the spherical radius. Assuming that there is a
   regular center ($r=0$), we put there $l=0$ without loss of generality,
   and we will classify the configurations under study by the limiting value
   of $r(l)$ (infinite, finite, or zero) at the largest or infinite
   values of $l$.

   In the general monopole case, the regular center requirement leads to the
   following boundary conditions for \eqs
   (\ref{phi''})--(\ref{int}) at $l=0$:
\beq
    \phi (0) =\gamma'(0)  = r(0) =0, \qquad r' (0) = 1.
                                    \label{cond at l=0}
\eeq
   The system (\ref{phi''})--(\ref{int}) does not contain $\gamma$ but
   only its derivatives. It is convenient for numerical integration to work
   with \eqs (\ref{phi''})--(\ref{beta''}) solved with respect to the
   second-order derivatives and consider (\ref{int}) as their first integral.

   We thus have four boundary conditions (\ref{cond at l=0}) for the
   (effectively) fourth-order set of equations. It might seem that we must
   obtain a unique solution. However, this is not the case since $l = 0$,
   being a singular point of the spherical coordinate system (not to be
   confused with a space-time curvature singularity), is also a singular
   point of our set of equations. As a result, our set of equations admits
   an additional freedom of choosing $\phi' (0)$; or, instead, we may use
   the requirement of global regularity to obtain a unique solution.

\subsection*{Infinite extra dimensions}

   If the solution is defined in the interval $0 \leq l < \infty $, then the
   lacking boundary condition can be taken as $\phi \to \const$ as $l \to
   \infty$, or
\beq
    \phi' (\infty)  =0.                     \label{Fi'(inf)=0}
\eeq
   In general, when the scalar field starts from a maximum of the potential
   and ends at a finite value of $\phi$, the five boundary conditions
   (\ref{cond at l=0}) and (\ref{Fi'(inf)=0}) uniquely determine a
   nontrivial solution to our field equations. Its existence determines a
   certain area in the space of parameters of the problem without {\it a
   priori\/} fine tuning. An asymptotic analysis at $l\to \infty $ shows
   that in this general case $\beta (l)$ is a linearly growing function as
   $l\to \infty$, and \ $r' (\infty) \geq 0$.

   $r' (\infty)=0$ is the special case when $r$ tends to a finite constant
   at $l\to \infty $. The solution then terminates at a slope of the
   potential rather than its minimum. The supplementary condition $r'(\infty)
   = 0$ seems to require a fine tuning relation between the free parameters
   of the problem. But we shall see that it is not quite so. An analysis
   of solutions with $\phi =\phi_0 =\const$ shows that there is also an
   area in the parameter space where the condition (\ref{Fi'(inf)=0}) is
   fulfilled automatically, and solutions with $r' (l) \to 0 $ as $l\to
   \infty $ exist without any fine tuning.

\subsection*{Two centers}

   It can happen that the integral curves of a solution terminate at some
   finite value $l_c$ with $r(l_c)=0$ and $\phi(l_c)=0$, which is one more
   center. Of interest for us are configurations in which this second
   center $l = l_c$ is also regular. Then the same four conditions
   (\ref{cond at l=0}) should hold at $l=l_c$ too. Two of them can be
   fulfilled by choosing the values of $\phi' (0)$ and $l_c$. Two others
   can only be fulfilled by a proper choice of free (input) parameters
   of the problem, e.g., those of the potential (if any) and the
   cosmological constant.

   In the special case of symmetry between the centers\footnote
{\eqs (\ref{phi''})--(\ref{int}) are invariant under translations $l \to
l+l_0$ and under reflections $l_0+l \to l_0-l, \phi \to  -\phi$. This
invariance gives rise to existence of regular solutions with two centers,
which are symmetric against the middle point \cite{we-05}.

Furthermore, a solution with two regular centers defined in the interval
($0, l_c$) can be symmetrically extended to the next interval ($l_c,
2l_c$) and further on, thus leading to a periodic solution for $l\in \R$
\cite{Brihaye}. The metric remains regular everywhere; however, the points
of contacts ($0,\ \pm l_c,\ \pm 2l_c, ...)$ are geometrically ambiguous:
each of them belongs to two adjacent manifolds simultaneously.

If one still believes in the reality of such systems, one can note
that the spectrum of a low-energy particle in a perfectly periodic
potential has a conductivity zone, allowing free propagation and
thus making the extra dimension observable in principle. However,
if the conductivity zone is very narrow, then even small
perturbations should lead to localization of a particle. This
interesting new possibility is worth a special consideration.},
   the input parameters are connected by only one fine-tuning relation. In
   this case, the boundary conditions at the second center are satisfied
   automatically, and the existence of a regular solution is provided by
   three conditions of smoothness at the middle (equator) point $l_{\rm eq}
   = l_{c}/2$:
\[
    \phi ' (l_{\rm eq}) = \gamma ' (l_{\rm eq}) = r' (l_{\rm eq}) = 0.
\]
   Two of these three conditions determine the values of $l_{c}$ and
   $\phi ' (0)$, and the remaining one requires fine tuning of the input
   parameters of the problem.

   The fine tuning could be avoided at the expense of admitting conical
   singularities (in the string case) at the two centers. For symmetric
   solutions the three smoothness conditions at the equator could be
   satisfied by choosing $\phi' (0)$, $r'(0)$ and $l_{c}$. In the general
   case of nonsymmetric centers, the four conditions of the second center
   could be satisfied by appropriately choosing $\phi' (0)$, $r'(0)$,
   $r'(l_c)$ and $l_{c}$.
\newpage

\subsection{String equations}

   In the string case $d_1 =1$ to be considered here in detail,
   the field equations take the form
\bear
    \phi'' +( d_0 \gamma ' + \beta'{}^2) \phi '
    - \e^{-2\beta }\phi \eql \frac{\d V}{\d \phi},      \label{eq-phi}
\\
     \gamma'' +d_0 \gamma '{}^2 +\beta' \gamma'
        \eql -\frac{2\kappa^2}{d_0}V,               \label{eq-gamma}
\\
    \beta'' +d_0 \beta' \gamma' + \beta '{}^2
    \eql -(\kappa^2 \phi^2 ) \e^{-2\beta } - \frac{2\kappa^2 }{d_0}V,
                                        \label{eq-beta}
\\
    2d_0 \gamma'\beta ' + d_0 (d_0 -1) \gamma'{}^2
     \eql \kappa^2 ( \phi'{}^2 -2V) -\kappa^2 \phi^2 \e^{-2\beta}.
                                        \label{int-s}
\ear

   For numerical examples, we will use the so-called Mexican hat potential
\bear
    V = \frac{1}{4} \lambda_0\eta^4                          \label{hat}
      \biggl[\eps + \biggl( 1-\frac{\phi^2}{\eta^2}\biggr)^2 \biggr].
\ear
   The parameter $\eps$ plays the role of a cosmological constant added to
   the conventional Mexican hat potential (the ``hat'' is thus moved up or
   down). To pass to dimensionless quantities, we put $\lambda_0 \eta^2 =1$,
   so that lengths will be measured in units of $1/(\sqrt{\lambda_0}\eta)$
   which in many cases has the meaning of the string core radius\footnote
{From the very beginning, we put $c = \hbar =1$, so that all quantities are
 measured in appropriate powers of length $[\ell]$. Then some relevant
 dimensionalities are $[V] = [\ell^{-D}]$,
 $[\phi^2] = [\eta^2] = [\kappa^{-2}] = [\ell^{2-D}]$. }.

   The potential then takes the form
\beq                                                            \label{hat'}
    V = \frac{1}{4} \eta^2 [\eps + (1-f^2)^2],
            \cm   f := \frac{\phi}{\eta}.
\eeq
   The remaining free parameters that control the system behavior are
\beq                                                    \label{param}
    d_0, \quad \eps, \quad {\rm and} \quad \Gamma := \kappa^2 \eta^2 .
\eeq
   We will use $d_0=4$ in computations. The parameter $\Gamma$ characterizes
   the gravitational field strength.

   In the further description of different regular solutions to our
   field equations, we begin with the simplest solutions in which
   $\phi = \phi_* = \const$. They are not string solutions but are helpful
   for comparison.

\section{Solutions with $\phi =\const$}

    If $\phi = \phi_* = \const$, we are actually dealing with vacuum field
    equations for the metric (\ref{Metric}) with the cosmological constant
    $\Lambda = \kappa^2 V(\phi_*)$.

    The scalar field equation (\ref{eq-phi}) reduces to
\beq                                                         \label{fi_0}
    \phi_*/r^2 = -V_\phi (\phi_*), \cm V_\phi := dV/d\phi.
\eeq
    This leads to two kinds of solutions: one exists in the case
    $V_\phi(\phi_*) = 0$, $\phi_*=0$, the other corresponds to
    $V_\phi (\phi_*) \ne 0$ (i.e., $\phi$ is ``frozen'' on a slope of the
    potential), and in this case we should put $r = \const$.

    For the potential (\ref{hat}), \eq (\ref{fi_0}) gives
\beq
       f_* ( r^{-2} - 1 +f_*^2 ) =0,                    \label{f0=0}
\eeq
    so that either $f_* = \phi_*/\eta = 0$ or $f_*^2 = 1 - 1/r^2$.

\subsection{Solutions with $\phi \equiv 0$}

    The trivial regular solutions with the order parameter $\phi\equiv 0$
    describe configurations with a higher symmetry, which can become
    spontaneously broken into a structure with a topological defect.

    In this case the metric obeys the equations
\bear
    \gamma'' +\gamma'( d_0 \gamma' +\beta ') \eql - 8\Lambda/d_0,
\nnv
    \beta '' +\beta' ( d_0 \gamma' +\beta') \eql -8\Lambda/d_0,
\nnv
     (d_0 -1) \gamma'{}^2 +2\gamma' \beta' \eql -8\Lambda/d_0.\label{eq-f0}
\ear
    Excluding $\beta' $, we get the equation for $\gamma$
\beq
      \gamma '' +\frac{1}{2}( d_0 +1) \gamma'{}^2 +\frac{4\Lambda}{d_0}=0
                                                \label{gam'',f=0}
\eeq
    Its solutions are different for positive and negative $\Lambda $.
    For $\Lambda >0$ we get (requiring $\gamma(0) = \gamma'(0) = 0$ --- a
    regular center at $l=0$)
\beq                                                         \label{gamma_>}
    \e^{(d_0+1) \gamma} = \cos^2 (\lambda_1 l), \cm
                \lambda_1 = \sqrt{2\Lambda \frac{d_0 + 1}{d_0}}.
\eeq
    The last equation (\ref{eq-f0}) then gives for $r = \e^\beta$:
\beq
      r^2 = r_0^2 \e^{-(d_0-1)\gamma} \sin^2 (\lambda_1 l),
            \cm r_0 = \const,                             \label{r_>}
\eeq
    where, choosing $r_0 = 1/\lambda_1$, we can satisfy the regularity
    condition $r'(0)=1$. We thus have a configuration with a regular center
    but with  a singularity $\e^\gamma \to 0$ and $r \to \infty$ as
    $l \to \pi/(2\lambda_1)$.

    For the potential (\ref{hat}), this case corresponds to $\eps > -1$.

    For $\Lambda <0 $, corresponding to $\eps < -1$, we have, instead of
    (\ref{gamma_>}) and (\ref{r_>}) \footnote
{For $d_0=4$ these formulae reduce to those found earlier in
\cite{Cline}.},
\bear \label{gamma_<}
      \e^{ \gamma} \eql \cosh^{2/(d_0+1)} (\lambda_2 l), \cm
            \lambda_2 = \sqrt{-2\Lambda \frac{d_0+1}{d_0}}.
\\
     r \eql r_0 \frac{\sinh (\lambda_2 l)}
            {[\cosh(\lambda_2 l)]^{(d_0-1)/(d_0+1)}},       \label{r_<}
\ear
    and again, choosing $r_0 = 1/\lambda_2$, we can satisfy the regularity
    condition $r'(0)=1$.

    So the configuration with unbroken symmetry is completely regular and
    extends from the regular center $l=0$ to $l=\infty$ where the warp
    factor $\e^{2\gamma}$ and the radius $r$ are infinitely growing functions.

\subsection{Solution with $\phi = \phi_* = \const \neq 0$}

    In this case, \eq (\ref{fi_0}) leads to a constant radius $r = r_*$,
    and \eq (\ref{eq-beta}) gives the relation
\beq
     \phi_*^2/r_*^2 =  -V(\phi_*)/d_0,
\eeq
    whence it follows $V(\phi_*) = \Lambda/\kappa^2 < 0$.

    For $\gamma$ we get from (\ref{int-s})
\beq                                                       \label{gam-cy}
      d_0^2\gamma'{}^2 = - 2\Lambda, \cm
                \e^{d_0\gamma} = \e^{\pm \sqrt{-2\Lambda} l},
\eeq
    and the coordinate range is $l \in \R$. In particular, for the potential
    (\ref{hat}) we have from (\ref{f0=0}) and (\ref{eq-beta})
\bear
    r \eql 1/ \sqrt{1-f_*^2} = \const,                      \label{r-cy}
\\                                                          \label{eps-cy}
    \eps \eql - 2d_0 (1 - f_*^2) f_*^2 - (1-f_*^2)^2 < 0,
\\                                                          \label{Lam-cy}
    \Lambda \eql \kappa^2 V(\phi_*)
            = -\Half\kappa^2 \eta^2 d_0 f_*^2 (1-f_*^2)].
\ear
    The configurations with $\phi = \const \neq 0$ and $r = \const$ are
    regular but evidently cannot be interpreted in terms of a brane world.
    They only provide the asymptotic behavior of the ``tube'' solutions
    presented below.

\begin{table*}
\caption{Classification of regular ($d_0+2$)-dimensional solutions
	for arbitrary $V(\phi)$ by the types of asymptotic behavior at the
	largest or infinite $l$. The columns denoted $r$, $\phi,$ and
	$\gamma$ show their final values. Attraction or repulsion is meant
	with respect to the center.  The symbol $\eta$ means a minimum of
        $V(\phi)$.}

\begin{center}                          \tabcolsep 10pt
\begin{tabular}{|c|c|c|c|c|c|l|}
\hline
 Notation   & $l$ range & $r$  &  $\phi$   &  $V(\phi)$ & $\gamma$  &
                                         Asymptotic type $\wide$\\
\hline
       &&&&&&\\[-10pt]
  A0  &  $\R_+$ & $\infty$ & $\equiv 0$  & $V(0) < 0$ & $\infty$ &
                                 AdS, attraction \\
  A1  &  $\R_+$ & $\infty$ & $\eta$   & $V(\eta) < 0$ & $\infty$ &
                                 AdS, attraction \\
  A2  &  $\R_+$ & $\infty$ & 0        &    0    & $\infty$
                        & power-law, attraction \\
       &&&&&&\\[-10pt]
\hline
       &&&&&&\\[-10pt]
  B0  &  $\R\  $  & $\equiv r_*$ & $\equiv \phi_* \ne 0$
                    & $V_*<0$ & $\pm \infty$
                                 & horizon at one end \\
  B1  & $\R_+$ & $r_*$ & $\phi_* \ne 0$ &   $V_*<0 $ & $-\infty$
                                 & horizon, repulsion \\
  B2  & $\R_+$ & $r_*$ & $\phi_* \ne 0$ &   $V_*<0 $ &  $\infty$
                             & attracting tube \\
       &&&&&&\\[-10pt]
\hline
  C  & $(0,\,l_c)$  &  0  &  0 &  $V(0)>0$ &  const & second  center$\wide$\\
\hline
\end{tabular}
\end{center}
\end{table*}

\section{Regular string solutions.
    Classification and map in the $(\Gamma,\ \eps)$ plane}

   The possible types of regular solutions to our field equations have been
   classified in Refs.\,\cite{we-05, we-06}. Table 1 represents this
   classification for the string case $d_1 = 1$. Compared to \cite{we-05},
   this table omits the solutions existing only for $d_1 > 1,$ but
   additionally includes the solutions (labelled A0 and B0) from \sect 3.

\begin{figure}[h]
\hspace{3cm}
\includegraphics{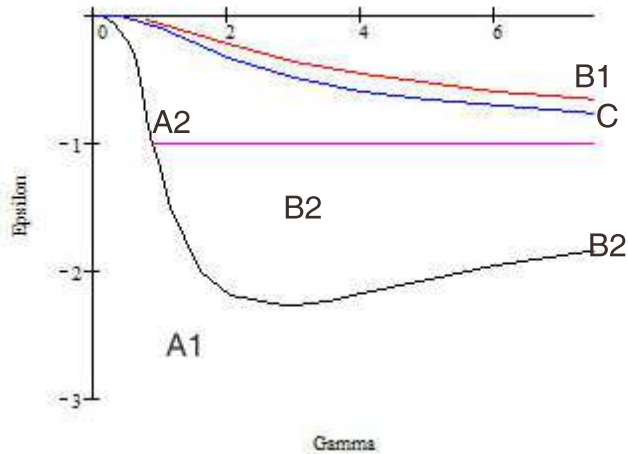}
\caption{Location of solutions in the plane of parameters $(\Gamma,\eps)$.
    The black line $\eps_* (\Gamma)$ is the upper boundary of the area of
    class A1 solutions with $r\to \infty$ as $l\to \infty $. The black line
    itself and the whole area $-1 > \eps \geq \eps_*(\Gamma)$ correspond to
    class B2 solutions with $r(\infty)= \const$. Fine-tuned solutions with
    two symmetric regular centers (class C) are located along the blue line
    $\eps_1 (\Gamma)$. The class B1 of fine-tuned solutions with a horizon
    is represented by the red line $\eps_h (\Gamma)$.}  \label{map2}
\end{figure}

   In what follows, we will deal with the potential (\ref{hat'}). For this
   potential, Fig.\,\ref{map2} presents the location of different regular
   string solutions in the plane of parameters $(\Gamma,\ \eps)$. The map
   shows solutions with the $\phi$ field having a constant sign. Those with
   alternate signs of $\phi$ will be discussed in \sect 6. There
   are no regular string solutions at $\eps > 0.$

   In Fig.\,\ref{map2}, the black line $\eps_* (\Gamma)$ is the upper
   boundary of the area of class A1 solutions with $r\to \infty$ as $l\to
   \infty $. The points on the black line itself and in the whole area $-1 >
   \eps \geq \eps_*(\Gamma)$ correspond to class B2 solutions. Fine-tuned
   solutions with two symmetric regular centers (see Fig.\,\ref{2c}) are
   located along the blue line. The class of fine-tuned solutions with a
   horizon (B1) is represented by the red line.

   Let us briefly describe the classes of solutions presented in the map,
   postponing the derivation of some important details of the curves to the
   subsequent sections.

\subsection*{A: Configurations with infinite $r$}

   From Table 1, one can notice a close similarity between the vacuum
   solutions A0 (where $\phi \equiv 0$) and A1. In fact, the gravitational
   field in both cases is mainly governed by the (negative) cosmological
   constant. In the limiting case $|\eps| \gg 1, \eps < 0$ the role of the
   symmetry breaking potential $V(\phi)$ is negligible. Then, as follows
   from (\ref{eq-gamma}), $\gamma'$ is always positive, the warp factor
   $\e^{2\gamma}$ grows, and gravity is attractive towards $l=0$. These
   solutions with $r\to \infty $ at large $l$ exist without any fine tuning.

   As $|\eps|$ decreases, the potential $V(\phi)$ becomes more and more
   important, and at $\eps$ approaching some $\eps_*(\Gamma)$ the derivative
   $r' \to 0$, so that class A1 solutions pass over to asymptotically
   cylindrical fine-tuned class B2 solutions.

   The main features of class A1 solutions at large $l$ are:

    --- the scalar field $\phi(l)$ tends to a minimum of $V(\phi)$;

    --- the quantities $\gamma'(l)$ and $\beta'(l)$ tend to the same finite
    positive constant, so that $\e^{\gamma(l)} \sim r(l)$ grow exponentially.

\begin{figure}[h]
 \hspace{-0,5cm}
 \includegraphics{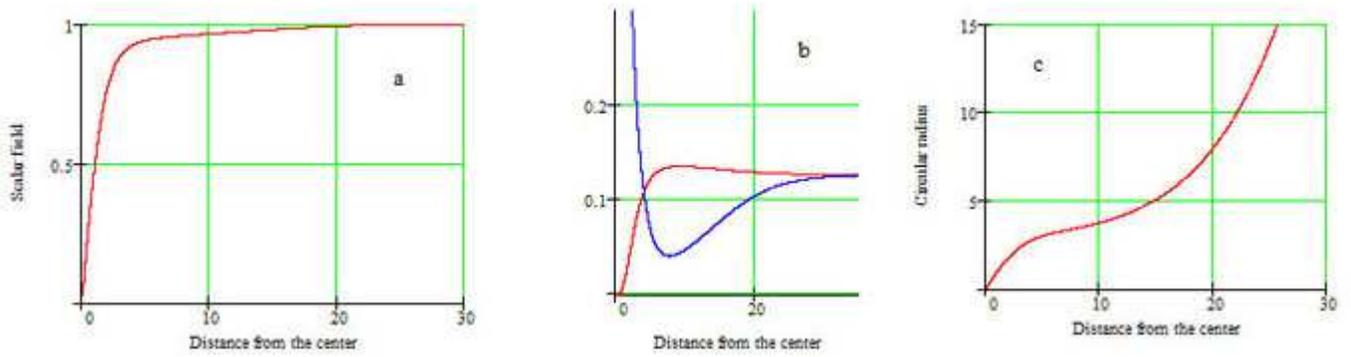}
\caption {A1 solution with the parameters: $\Gamma =0.7$, $\eps =-0.9$.
    (a) --- scalar field, \ (b, blue) --- $\beta'(l)$, \ (b, red) ---
    $\gamma ' (l)$, \ (c) --- $r(l)$.\label{rto inf} }
\end{figure}

    An example of an A1 solution, found numerically, is presented in
    Fig.\,\ref{rto inf}.

    As to the A2 regime, this is an exceptional solution corresponding to
    the condition $V(0) = 0$, hence $\eps =-1$. In this case \cite{we-06}
    a regular integral curve, starting at $l = 0$ with $\phi = 0$, finishes
    again with $\phi \to 0$ as $l \to \infty$. The large $l$ behavior
    of $r$ and
    the warp factor $\e^{2\gamma}$ in the resulting regular solutions is
\[
    r \approx  l, \cm  \e^{d_0 \gamma} \sim l.
\]

\subsection*{B: Asymptotically cylindrical configurations}

    For these ``tube'' solutions, it is easy to verify that \eqs
    (\ref{gam-cy})--(\ref{eps-cy}) hold at large $l$ and, in particular,
    $f \to f_* = \const$ where $0 < f_*^2 < 1$.

\begin{figure}[h]
\hspace{4cm}
\includegraphics{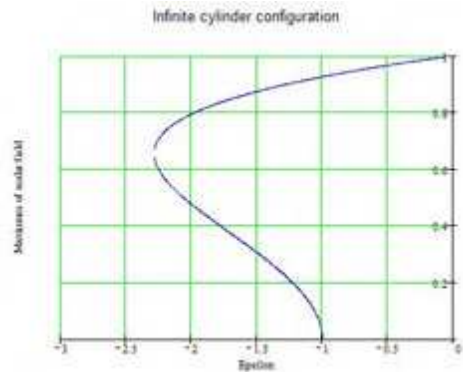}
\caption{The dependence $\eps (f_*)$ (\ref{eps-cy}) in the
	infinite cylinder case}
                        \label{eps(f)}
\end{figure}

    The vacuum solution B0, with $r \equiv  r_*$, actually interpolates
    between the cylindrical asymptotics of B1 ($\e^{\gamma} \to 0$, a
    double horizon) and B2 ($\e^{\gamma} \to \infty$ at $l \to \infty$,
    i.e., gravitational attraction towards the center $l=0$).

    \eq (\ref{eps-cy}) does not contain $\kappa$ and allows finding the
    range of the input parameter $\eps$ for which ``tube'' solutions are
    possible. The dependence $\eps(f_*)$ is shown in Fig.\,\ref{eps(f)}.
    By \eq (\ref{eps-cy}),
\beq
    \eps >\eps_{\min} = -1 - \frac{(d_0 -1)^2}{2d_0 -1}  \label{Eps min}
\eeq
    For $d_0 =4$, $\eps_{\min } = -16/7 = -2.2857...$.

    It also follows from (\ref{eps-cy}) and Fig.\,\ref{eps(f)} that in the
    range $-1 > \eps >\eps_{\min }$ there are two branches of the inverse
    function $f_*(\eps)$. In the range $0\geq \eps > -1$ there is only one
    branch. Other limiting values are expressed in terms of $f_*$ by
    (\ref{gam-cy})--(\ref{Lam-cy}), and
\beq
    \gamma' \to \kappa \eta \sqrt{d_0 f_*^2 (1-f_*^2)}.
\eeq

    Type B2 configurations occupy a whole area in the ($\eps,\,\Gamma$)
    parameter plane, whereas B1 solutions require fine tuning and are located
    on the red line in the map.

\subsection*{C: Configurations with two centers}

    As was argued above, type C solutions can be symmetric and asymmetric
    with respect to reflections $l \to l_c-l$. Symmetric solutions require
    one fine-tuning relation, which corresponds to particular curves in the
    ($\eps,\,\Gamma$) plane. The curve describing solutions with a constant
    sign of $\phi$ is presented in Fig.\,\ref{map2} (the blue line). Other
    symmetric configurations will be discussed below. Asymmetric solutions
    can only appear at discrete points in the parameter plane, and we will
    not mention them any more.

\section{``Tube'' solutions: location in the parameter plane}

\begin{figure}[h]
\hspace{2cm}
\includegraphics{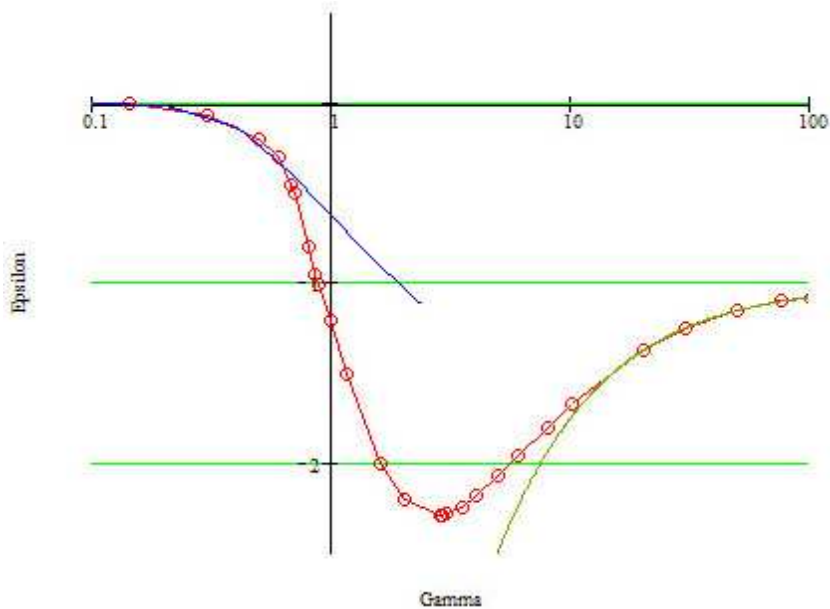}
\caption{The upper boundary $\eps_*(\Gamma)$ of class A1 solutions with
infinite $r$. The points forming the red line are found numerically.
The asymptotic dependences for $\Gamma\ll 1$ (\ref{weak!}) and $\Gamma\gg 1$
(\ref{strong!}) are shown by the blue and brown curves, respectively.
}                                                     \label{map}
\end{figure}

    The upper boundary $\eps_*(\Gamma)$ of type A solutions, found
    numerically point by point for $d_0=4$, is presented in Fig.\,\ref{map}
    by the red line and the circles. Any point in the area
    $\eps < \eps_*(\Gamma)$, $0 < \Gamma < \infty$ corresponds to a type A
    solution with $f$ monotonically growing from zero at the center to
    unity. The function $\eps_* (\Gamma)$ decreases from zero at $\Gamma
    =0$ to a minimum with $\eps =\eps_{\min }$ according to \eq (\ref{Eps
    min}) and then grows tending to $-1$ as $\Gamma \to \infty$.

    In the range $0 > \eps > -1$, the ``tube'' (fine-tuned) solutions exist
    only precisely on the line $\eps_*(\Gamma)$, which comprises a border
    between A and C type solutions.

    In the range $-1 > \eps > \eps_* (\Gamma)$, there are cylindrical
    solutions without fine tuning. This area is located between the zones of
    A and C type solutions.

    In the limits of weak and strong gravitational fields (small and
    large $\Gamma$, respectively), a numerical analysis of the field
    equations is hindered, and we have derived the function $\eps_*(\Gamma)$
    analytically. The blue line in Fig.\,\ref{map} corresponds to
    $\Gamma \ll 1$, the brown one to $\Gamma \gg 1$.

\subsection{Strong gravity: $\Gamma \gg 1$}

    As follows from numerical analysis, the scalar field in ``tube''
    solutions is small in this case, and the potential can be expanded in
    a series:
    $V(\phi) \approx V_0 +\frac{1}{2}V''_0\phi^2$,
    $dV/d\phi \approx V''_0\phi$.
    The problem is completely determined by the two constants $V_0 =V(0)$
    and $V''_0 = (d^2 V/d\phi^2)_{\phi=0}$. Let us introduce a new parameter
    $\lambda $ and a new function $\psi $:
\beq
    \lambda =4\kappa^2 V_0 ,\qquad \psi =\kappa \phi .  \label{lambda=}
\eeq
    For the Mexican hat potential (\ref{hat}), $V_0 =\eta^2(\eps+1)/4$ and
    $V''_0 =-1$, so that
\beq
    \lambda = \Gamma ( \eps +1)                \label{lambda-hat}
\eeq
    (recall that $\Gamma:= \kappa^2 \eta^2$). \eqs
    (\ref{eq-phi})--(\ref{eq-beta}) take the form
\bear
    \gamma'' +\gamma' (d_0 \gamma'+\beta') &=&
            -\lambda/(2d_0) + \psi^2/d_0, \label{strong1}
\\
    \beta '' +\beta ' (d_0 \gamma' +\beta') &=&
        \psi^2 (1/d_0 - \e^{-2\beta }) - \lambda/(2d_0),
                            \label{strong2}
\\
    \psi'' +\psi' ( d_0 \gamma ' +\beta') &=&
                \psi (\e^{-2\beta } -1)  \label{strong3}
\ear
   and depend on only one dimensionless parameter $\lambda $. The boundary
   conditions are
\[
    \gamma (0)  =\gamma'(0) =\psi (0) =0,\qquad (\e^{\beta})'\big|_{l=0}=1,
    \qquad  \beta' (\infty)  =\psi' (\infty) =0.
\]
   The scalar field equation (\ref{strong3}) is homogeneous with respect
   to $\psi $ and looks linear; however, the system as a whole is nonlinear,
   and we have a nonlinear eigenvalue problem. The parameter $d_0$ being
   fixed, there is only one dimensionless parameter $\lambda $, whose ground
   state eigenvalue is expected to be of the order of unity. Then according
   to (\ref{lambda-hat}) for $\Gamma \gg 1$, the parameter $\eps $ is
   very close to $-1$, and $f_{0}$ in (\ref{eps-cy}) is $\sim \Gamma^{-1}
   \ll 1$. From (\ref{strong3}) it follows that $r \equiv  \e^{\beta}\to 1$
   at large $l$. Nontrivial solutions exist for discrete values of
   $\lambda $, one of which, corresponding to a monotonically growing $\psi
   (l)$, is found numerically:
\[
    \lambda =-7.433...,\qquad d_0 =4.
\]
   The asymptotic dependence $\eps_* ( \Gamma ) $ for $\Gamma \gg 1$
\beq    \label{strong!}
    \eps = - 1 + \lambda/\Gamma ,\qquad \Gamma \gg 1
\eeq
   is presented in Fig.\,\ref{map} by the brown curve.

\subsection{Weak gravity: $\Gamma \ll 1$}

   The case $\Gamma \ll 1$ is more complicated. A numerical computation
   shows (and it is verified analytically) that $| \eps|$ is exponentially
   small as $\Gamma \to 0$. From (\ref{eps-cy}) we see that
   $1-f_{0}^2 \approx - \eps/(2d_0) \ll 1$, and the limiting value of
   the circular radius (\ref{r-cy}),
    $r_* \approx \sqrt{2d_0/| \eps}$
   is very large as compared with the ``core'' radius $ \sim 1$. The
   equations simplify differently in the two cases $r\ll r_*$ and $r\gg 1$.
   The solutions should coincide in the intermediate region $1\ll r\ll r_*$.

   It is convenient, for $\Gamma \ll 1$, to rewrite the field equations
   in terms of $r=\e^{\beta }$:
\bear
    \gamma'' \eql -\gamma ' \left( d_0 \gamma ' -\frac{r' }{r}\right)
    -\frac{\Gamma  }{2d_0 }\left[ \eps +( 1-f^2)^2 \right],
                                \label{weak1}
\\                                                      \label{weak2}
    r'' \eql \frac{ (d_0 -1) d_0 }{2} \gamma'{}^2\, r
    -\frac{\Gamma}{2} \frac{f^2 }{r} - \frac{\Gamma}{2}f'{}^2 r
            - \frac{\Gamma}{4}\left[ \eps + (1-f^2)^2 \right] r,
\\
    f'' \eql -f' \left( d_0 \gamma'
    -\frac{r'}{r}\right) + \frac{f}{r^2} -f (1-f^2).  \label{weak3}
\ear

   For $r\ll r_*$ we see that $\gamma ' \sim \Gamma \ll 1$, and
   the term with $\gamma'{}^2 $ in (\ref{weak2}) can be neglected. In the
   vicinity of the center, in the terms $\sim \Gamma  $, we can set
   $r=l$ and omit $\eps $. Then \eq (\ref{weak2}) reduces to
\[
    r'' = -\Gamma \frac{f_0^2 }{l}-\frac{\Gamma}{2d_0 }
    (1-f_0^2)^2 l ,\cm   \Gamma  \ll 1,\quad  l \ll r_*,
\]
   where $f_0 $ is the solution of (\ref{weak3}) with $\gamma ' =0$, \
   $r=l$,  and the boundary conditions $f(0)  =0$, \ $f(\infty)  =1$.
   With $r'(0) =1$, after integration we get
\[
    r' = 1 + \Gamma  \left[ -f_0^2 \ln l
    +2\int_{0}^{l}dl\,f_0 f'_0 \ln l
        -\frac{1}{2d_0 }\int_{0}^{l}dl( 1-f_0^2)^2 l  \right].
\]
   The integrals quickly converge for $l\gg 1$, and in the
   intermediate region $1\ll  l \ll r_*$ we have
\bear
     r'{}^2 = 1 + 2\Gamma  \left( -\ln r+2J_2 -\frac{J_1 }{2d_0 }\right) ,
    \qquad \Gamma \ll 1,\ \quad 1\ll r\ll r_*,
                              \label{weak4}
\ear
   where the integrals $J_1 $ and $J_2 $ are found numerically:
\beq
    J_1 =\int_{0}^{\infty} l\,dl\, ( 1-f_0^2)^2 =1,  \label{weak5}
   \qquad
    J_2 =\int_{0}^{\infty }dl\, f_0 f'_0 \ln l \cong 0.2.
\eeq

\begin{figure}[]
\hspace{-0cm}
\includegraphics{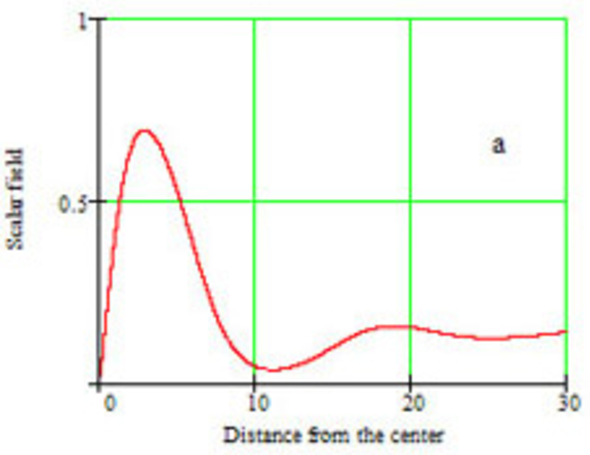}
\includegraphics{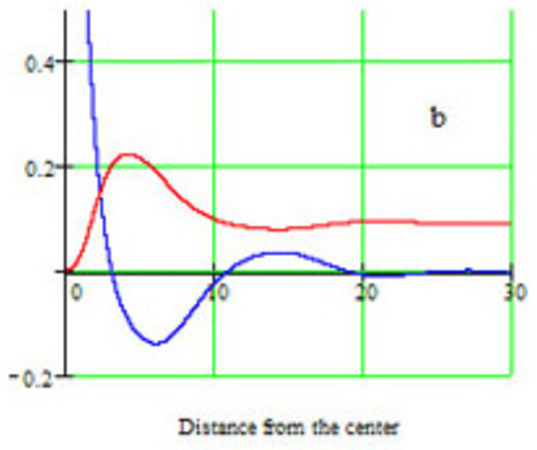}
\includegraphics{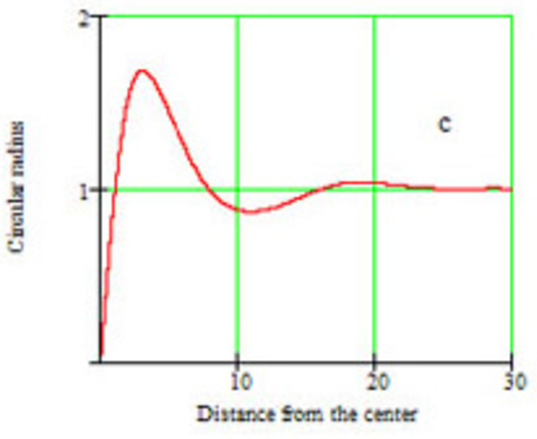}
\caption{Example of a type B2 solution
   for $\Gamma =2$ and $\eps =-1.1$. (a) shows the scalar field $f(l)$,
   (b) depicts $\gamma'(l) $ and $\beta'(l)$ by the red and
   blue curves, respectively, and (c) displays $r(l)$.
}                                       \label{r to rm}
\end{figure}

   In the region $r\gg 1$, \eq (\ref{weak2}) reduces to
\beq
    r'' =-\Gamma  \left(\frac{1}{r}+\frac{\eps }{2d_0 }r\right) ,
        \qquad \Gamma \ll 1,\ \ \ 1\ll r.
\eeq
   Taking into account that $r' =0$ at $r=r_*$, we find
\beq
    \frac{1}{2}r'{}^2 = \Gamma \left( \ln r_*-\ln r+\frac{\eps}
    {4d_0 }r_*^2 -\frac{\eps }{4d_0 }r^2 \right) ,\cm
        \Gamma  \ll 1,\ \ \ 1\ll r.         \label{weak6}
\eeq

   In the intermediate region $1 \ll r\ll r_*$ we have
\beq
    r'{}^2 =2\Gamma \left(\Half \ln \frac{2d_0 }{|\eps}
            -\ln r-\frac{1}{2}\right) ,\qquad
    \Gamma \ll 1,\ \ \ 1\ll r\ll r_*.       \label{weak7}
\eeq
   We have taken into account that $r_*=\sqrt{2d_0 /| \eps |}$.

   Comparing (\ref{weak4}) with (\ref{weak7}), we find the fine-tuning
   relation between $\eps $ and $\Gamma $ for asymptotically cylindrical
   configurations in the weak gravity limit:
\bear
    \eps &=& -2d_0 \exp {\left( -\frac{1}{\Gamma  }-4J_2
                    -1-\frac{J_1}{d_0}\right)}
        \cong -0.33\,d_0 \exp
            \left(-\frac{1}{\Gamma}+\frac{1}{d_0}\right) ,
\nnv
    \eps &\cong & -1.7\e^{-1/\Gamma}, \cm
          d_0 =4,\qquad \Gamma \ll 1.       \label{weak!}
\ear
   This asymptotic dependence $\eps_* (\Gamma)$ for $\Gamma \ll 1$
   is presented in Fig.\,\ref{map} by the blue curve.

\subsection{Solutions in the range $-1 > \eps > \eps_*(\Gamma)$}

   In the range $-1>\eps >\eps_*(\Gamma)$, there exist type B2 solutions
   ($ r\to r_* < \infty$ as $l\to \infty $) without any fine-tuning relation
   between the parameters $\eps$ and $\Gamma$. The asymptotic values of the
   scalar field ($f_*$) and the radius ($r_*$) at large $l$ do not depend on
   $\Gamma $:
\bear
     f_*^2 \eql \frac{d_0 -1-\sqrt{(d_0 -1)^2 +(\eps +1)(2d_0-1)}}{2d_0 -1},
\nnv
     r_*^2 \eql \frac{2d_0 -1}
        {d_0 +\sqrt{(d_0 -1)^2 +(\eps +1)(2d_0 -1)}},
\earn

   An example of such a solution is shown in Fig.\,\ref{r to rm}
   for $\Gamma =2$ and $\eps =-1.1$. The scalar field $f(l)$ is
   shown in Fig.\,\ref{r to rm}a, $\gamma'(l) $ and $\beta'(l)$ are red and
   blue curves in Fig.\,\ref{r to rm}b, and $r(l)$ is displayed in
   Fig.\,\ref{r to rm}c.

\subsection{Solutions with a horizon}

\begin{figure}[h]
\hspace{1cm}
\includegraphics{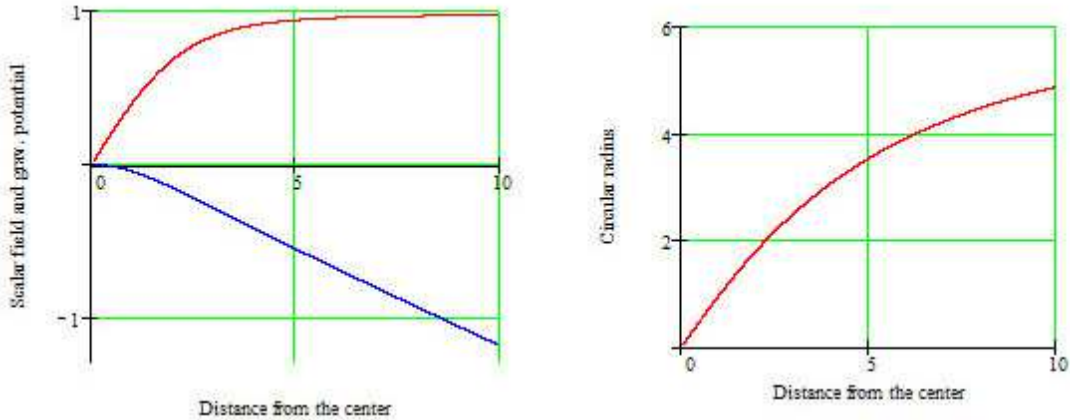}
\caption{A type B1 solution with a horizon for $\Gamma =2$ and
$\eps =-0.233846$. Left: the scalar field $\phi$ (red) and
the metric function $\gamma$ (blue). Right: the circular radius
$r$.} \label{hor}
\end{figure}

    As to type B1 configurations with a horizon, in which $\gamma(l)$
    linearly decreases as $l \to \infty$, their location, found numerically,
    is shown in Fig.\,\ref{map2} by the red line $\eps=\eps_{h}(\Gamma)$,
    corresponding to a certain fine-tuning relation. An example of such a
    regular solution, with the parameters $\Gamma =2$, \ $\eps =-0.233846$,
    is presented in Fig.\,\ref{hor}.

    The near-horizon metric has the asymptotic form
\beq \label{ds-B1}
      ds^2 = C^2\e^{-2hl} \eta\mn dx^\mu dx^\nu - dl^2 - r_*^2 d\Omega^2,
      		\cm   h=\gamma'(\infty).
\eeq
     The substitution $\e^{-h l} = \rho$ (converting $l=\infty$ to a finite
     coordinate value, $\rho=0$) brings the metric (\ref{ds-B1}) to the form
\beq \label{hor-B1}
      ds^2 = C^2 \rho^2 \eta\mn dx^\mu dx^\nu
                   - \frac{d\rho^2}{k^2 \rho^2} - r_*^2 d\Omega^2,
		   \cm  \rho \to 0.
\eeq
    Thus $\rho=0$ is a second-order Killing horizon in the 2-dimensional
    subspace parameterized by $t$ and $\rho$; the extra-dimensional circular
    radius squared remains positive. It is of the same nature as, e.g., the
    extreme \RN\ \bh\ horizon, or the \AdS\ horizon in the second
    Randall-Sundrum \bw\ model.  A peculiarity of the present horizon is
    that the spatial part of the metric, which at large $l$ takes the form
    $\rho^2 (d\vec x)^2$, is {\it degenerate\/} at $\rho=0$.  The volume of
    the $d_0$-dimensional spacetime vanishes as $l \to \infty$. And it will
    remain degenerate even if we pass to Kruskal-like coordinates in the
    ($t,\rho$) subspace. But the $D$-dimensional curvature is finite there,
    indicating that a transition to negative values of $\rho$ (where the old
    coordinate $l$ no longer works) is meaningful.

    Thus solutions with strings (and/or monopoles) in extra dimensions may
    contain second-order horizons, and the degenerate nature of the spatial
    metric at the horizon does not lead to a curvature singularity;
    moreover, the metric may be continued in a Kruskal-like manner. However,
    the zero volume of the spatial section makes the density of any
    additional (test) matter infinite at $\rho=0$. To consider these
    solutions as describing viable configurations, one needs to take into
    account the back-reaction of ordinary matter. It will evidently destroy
    such a configuration.

\section{Solutions with two regular centers: location in the parameter plane}

   Symmetric type C solutions with two regular centers are located on the
   $(\eps, \Gamma)$ plane in the region $0 > \eps > -1$ to the right from
   the {\bf red} fine-tuning curve $\eps_* (\Gamma)$ in Fig.\,\ref{map} or,
   which is the same, to the right of the black line in the full map,
   Fig.\,\ref{map2}.  The solutions are fine-tuned, i.e., located along
   certain lines $\eps_N (\Gamma)$ in this region, where $N$ is the number
   of half-waves, $N-1$ is the number of knots (zeros) of the scalar field
   $f = \phi/\eta$. The point is that $f$, just as the radius $r$, is zero
   at both centers, but $f$ can change its sign. So there are several
   families of regular solutions with different numbers $N$ of half-waves,
   each family corresponding to a line $\eps_N (\Lambda)$ in the parameter
   plane. The blue curve in Fig.\,\ref{map2} depicts $\eps_1(\Lambda)$.

\subsection{Solutions without knots of the scalar field
            \label{without knots}           }

\def\q{{\rm eq}}

   In solutions where $f$ has a constant sign, all three functions
   $f(l)$, $r(l)$ and $\gamma (l)$ reach their extremum values at the
   equator $l=l_\q$.  Setting in the first integral (\ref{int-s})
\beq
    f' (l_\q) =r' (l_\q) =\gamma'(l_\q) =0,    \label{'eq}
\eeq
    we find a relation between $f(l_\q) =: f_\q$ and $r(l_\q) =: r_\q$:
\beq
      r_\q^2 = \frac{2f_\q^2}{|\eps| - ( 1-f_\q^2 )^2}.   \label{r-eq}
\eeq
    It is convenient to use (\ref{r-eq}) together with (\ref{'eq}) as
    boundary conditions and perform numerical integration from the equator
    to one of the centers. Then the three conditions $f(l_c) =0$, $r(l_c)
    =0$ and $r'(l_c) =1$ determine the values of $f_\q$ and $l_c$ and a
    fine-tuning relation $\eps = \eps_1(\Gamma)$.

\begin{figure}[h]
	\hspace{.5cm}
	\includegraphics{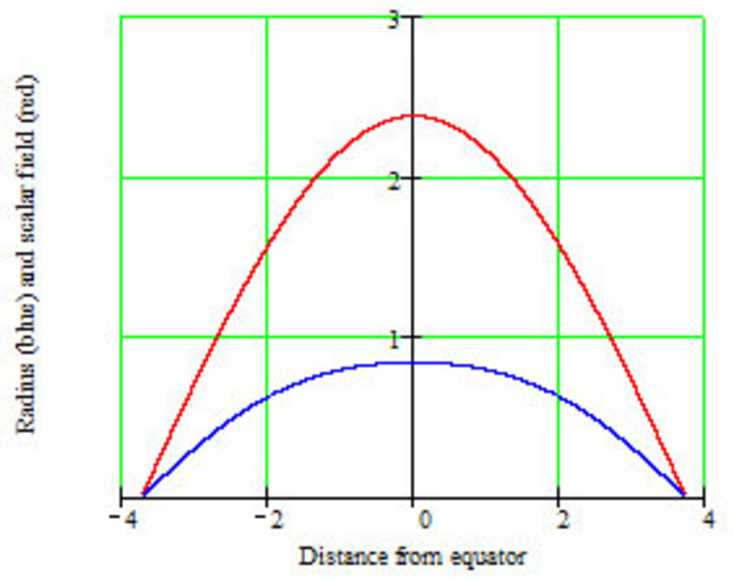}
	\hspace{1cm}
	\includegraphics{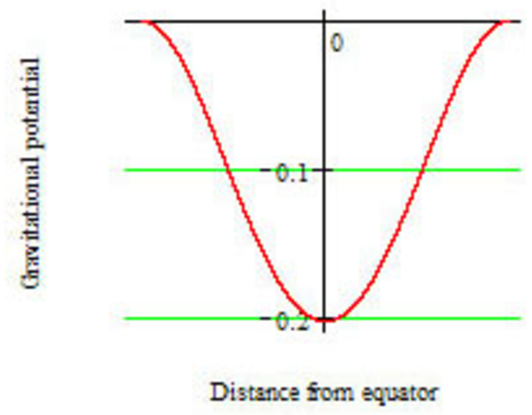}
\caption {Example of a configuration with two symmetric regular centers
without knots of the scalar field.  $\Gamma=2$, the fine-tuned
value  $\eps_1(2) =-0.3326...$. It belongs to the line $\eps=\eps_1(\Gamma)$.
Left: $r(l)$ --- red, $f(l)$ --- blue; right: $\gamma (l)$.
} \label{2c}
\end{figure}

    An example of a configuration with two regular centers is presented in
    Fig.\ref{2c} for $\Gamma=2$; the fine-tuned value of $\eps$ is
    $\eps_1(2) =-0.3326...$, it belongs to the line $\eps=\eps_1 (\Gamma)$.

    The (blue) curve $\eps = \eps_1(\Gamma)$ in Fig.\,\ref{map2} has been
    obtained numerically. For small and large values of $\Gamma$, this
    fine-tuning relation can be derived analytically.

\subsubsection*{$\eps_1 (\Gamma) $ for weak gravity, $\Gamma \ll 1$}

    This derivation repeats the one for \eq (\ref{weak!}). The main
    difference is that now we get from (\ref {r-eq}) the value of
    $r(l)$ at the equator
\beq
    r_\q = r(l_\q) =\sqrt{2/|\eps|}, \cm |\eps| \ll 1   \label{r(eq) weak}
\eeq
    and use it instead of $r_*$. Substituting (\ref{r(eq) weak})\ into (\ref{weak6}),
    we have instead of (\ref{weak7})
\[
    r'{}^2 = \Gamma \left( \ln \frac{2}{|\eps| }
    -2\ln r-\frac{1}{d_0 } - \frac{\eps}{2d_0 }r^2 \right) ,
                        \cm   \Gamma \ll 1,\ \ \  1\ll r.
\]
    In the intermediate region $1\ll r\ll r_m$ it should coincide with
    (\ref{weak4}). The resulting relation is
\beq
    \eps = -2\exp \left( -\frac{1}{\Gamma}
    - \frac{1-J_1}{d_0 }-4 J_2 \right) \cong -0.9\, \e^{-1/\Gamma},
        \cm \Gamma \ll 1.                   \label{eps1-weak,2c}
\eeq

\subsubsection*{$\eps_1 (\Gamma) $ for strong gravity, $\Gamma \gg 1$}

    Numerical integration shows that for $\Gamma \gg 1$ the scalar field
    $f$\ remains small in the whole interval between the centers, while
    $\psi =\kappa \eta f$ is of the order of unity. Introducing, as before,
    $\lambda $ (\ref{lambda-hat}) and taking into account that
    $f\ll 1$,\ we again find \eqs (\ref{strong1})---(\ref{strong3}).
    It is convenient to integrate them from the equator to a center and to
    use boundary conditions in the form
\beq
    \beta' (l_\q) =\psi' (l_\q) =\gamma' (l_\q) =0, \cm \psi (l_\q) =\psi _\q,
    \cm \beta (l_\q) =\frac{1}{2}\ln \frac{2\psi_\q^2 }{2\psi_\q^2 -\lambda }.
                        \label{boundary,strong}
\eeq
    The three parameters $\psi_\q$, $l_c-l_\q$ and $\lambda $ should be
    determined from the conditions
\[
    \psi (l_c) =0,\cm  r(l_c) =0,\cm      r'(l_c) =1.
\]
    Numerical integration results in $\lambda =1.9...$.

    Thus, for type C solutions, the desired fine-tuning relation in the
    strong gravity limit is
\beq
    \eps = - 1 + 1.9/\Gamma.     \label{eps1-strong,2c}
\eeq

\subsection{Odd and even scalar fields}

    In the above solutions, the scalar field $f(l)$ without knots is an
    even function.

    Since $f(l)$ may change its sign between the centers, there are two
    possibilities. If the number of knots of $f(l) $ is even, then $f(l)$ is
    an even function, reaching an extremum at the equator, and $f'(l_\q) =0$.
    On the contrary, $f(l)$ with an odd number of knots is an odd function:
    $f(l_\q) =0$, and $f'(l)$ is then an even function having an extremum at
    $l=l_\q$.

\begin{figure}[h]
\includegraphics{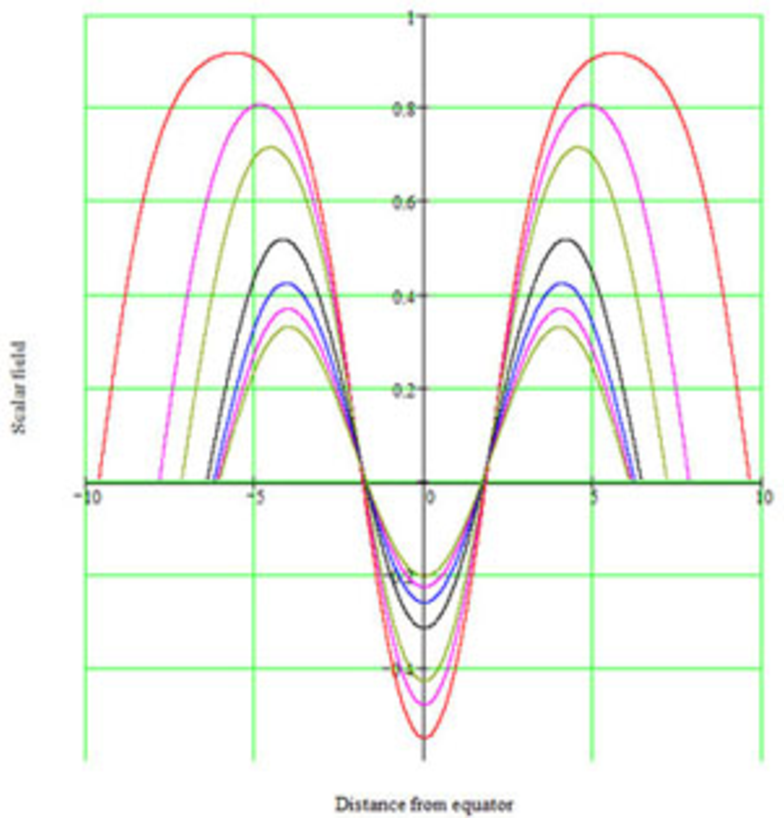}
\includegraphics{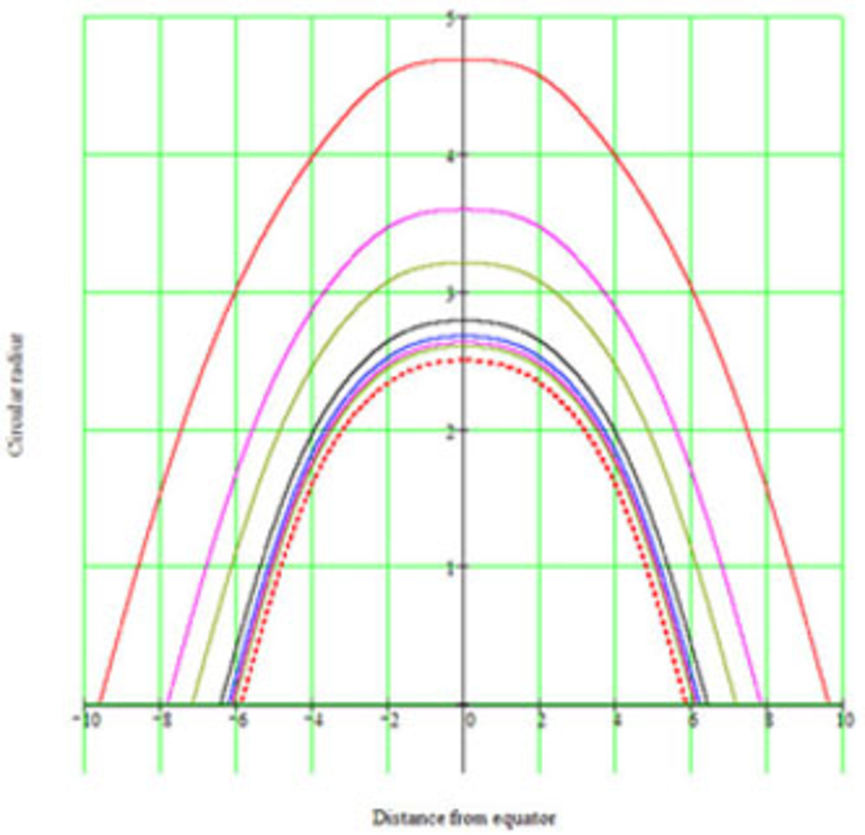}
\caption{Solutions with two symmetric regular centers with 2 knots
of $f(l)$ for $\Gamma =0.5$, 0.75, \ 1,\ 2,\ 3,\ 4 and 5 in the
order of decreasing amplitude. The corresponding fine-tuned values
of $\eps $ are $\ $-0.51275, -0.62765, -0.7019, -0.8365, -0.888,
-0.9146, and -0.93115. Left: $f(l)$. Right: $r(l)$; the additional
 dashed curve corresponds to the limit $\Gamma \to\infty $.
 } \label{2kn}
\end{figure}

\begin{figure}[]
\includegraphics{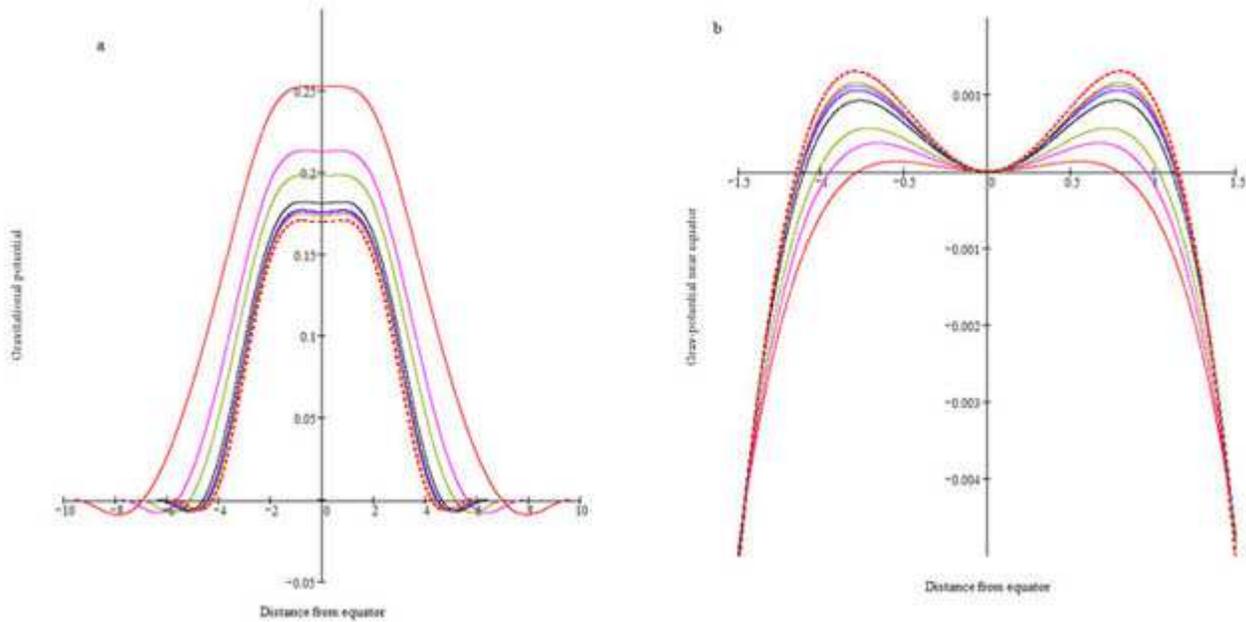}
\caption{The functions $\gamma (l) $ (a) in the whole range of $l$
and (b) in a close vicinity of the equator for the solutions with
     the same set of parameters as in Fig.\,\ref{2kn}.
     }
                                \label{2kn gamma(l)}
\end{figure}

\begin{figure}[]
\centering
\includegraphics{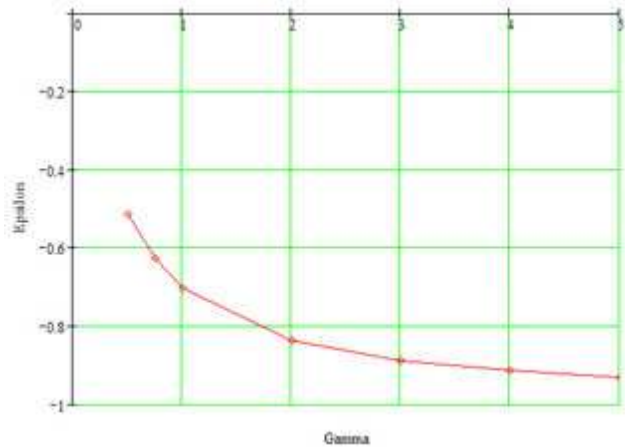}
\caption{The fine-tuning relation $\eps_3 (\Gamma) $ for solutions
     with two knots of $f(l)$ between the centers,
    displayed in Figs.\,\ref{2kn} and \ref{2kn gamma(l)}.
    }
                                    \label{2kn eps_2(Gamma)}
\end{figure}

    Numerical integration of \eqs (\ref{eq-phi})--(\ref{eq-beta}) in
    the case of an even number of knots can be performed with the same
    boundary conditions (\ref{boundary,strong}) as without knots. The results
    are displayed in Figs.\,\ref{2kn}--\ref{2kn eps_2(Gamma)}.

    Fig.\,\ref{2kn} shows a few solutions for the scalar field
    $f(l)$ with 2 knots and the corresponding functions $r(l)$.

   Fig.\,\ref{2kn gamma(l)} shows the function $\gamma (l)$ in the whole
   range of $l$ and (b) in a close vicinity of the equator for visual
   clarity, to demonstrate a minimum at the equator. Recall that $\gamma $
   enters into the equations only via $\gamma'$ and $\gamma''$, and so,
   without loss of generality, we have put $\gamma (l_\q) =0$ in
   Fig.\,\ref{2kn gamma(l)} (b). The larger is $\Gamma $,\ the deeper is the
   local minimum of $\gamma $\ at the equator. Altogether, the gravitational
   potential has three minima: one at the equator and two others near the
   regular centers.

   The fine-tuning relation $\eps_3(\Gamma)$ for solutions with two knots
   of $f(l)$, found numerically, is shown in Fig.\,\ref{2kn eps_2(Gamma)}.
   Each curve in Figs.\,\ref{2kn} and \ref{2kn gamma(l)} corresponds to a
   point on this curve.

\begin{figure}[h]
\centering
\includegraphics{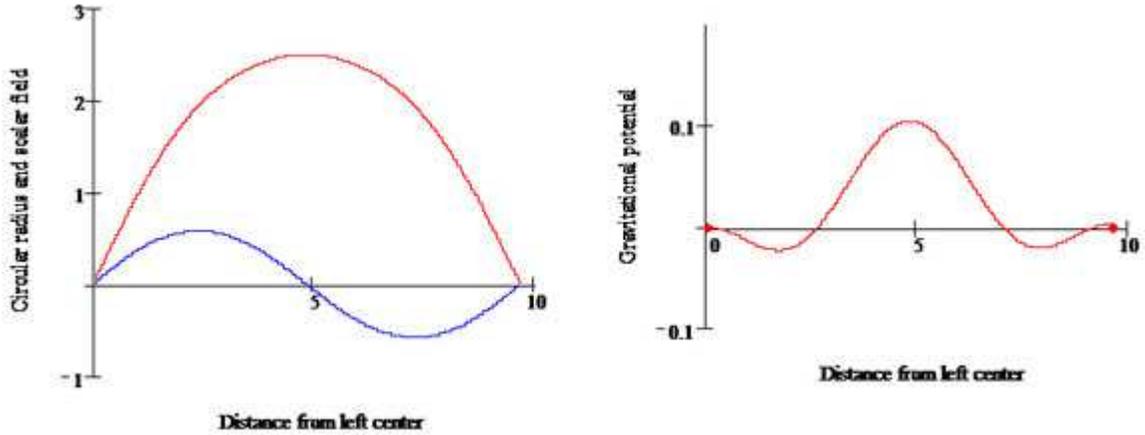}
\caption{Example of a solutions with two symmetric regular centers and
    an odd scalar field with one knot. $\Gamma =2$, the corresponding
    fine-tuned value $\eps_2(\Gamma) =-0.71$. The left graph shows $r(l)$
    (red) and $f(l)$ (blue); $\gamma (l)$ is depicted on the right graph.
   }
   \label{2c odd}
\end{figure}

    In solutions with an odd number of knots of $f(l)$, we have $f(l_\q)=0$,
    and from (\ref{int-s}) we obtain at the equator $f'{}^2 = (\eps +1)/2$.
    For numerical integration of \eqs (\ref{eq-phi})--(\ref {eq-beta}) in
    this case, it is convenient to use the following boundary conditions:
\beq
    \beta' (l_\q)=f(l_\q) =\gamma ' (l_\q) =0,
    \cm  \beta (l_\q) =\beta _\q,
    \cm  f'(l_\q) = \sqrt{(\eps +1)/2}.            \label{boundary-odd}
\eeq
    Then the three conditions $f(l_c) =0$, $r(l_c) =0$ and $r'(l_c) = 1$
    determine the values of $\beta_\q$ and $l_c-l_\q$ as well as the
    fine-tuning relation ($\eps= \eps_2(\Gamma)$ in the case of one knot).
    As an example, in  Fig.\,\ref{2c odd} we present a numerical solution
    with one knot for $\Gamma =2$, corresponding to $\eps =-0.71$. The left
    graph shows $r(l)$ (red) and $f(l)$ (blue). $\gamma (l)$ is depicted on
    the right graph. The function $\gamma (l)$ is symmetric relative to the
    equator and has two minima close to the centers.

    One can note that if we include into consideration configurations with
    angular deficits and excesses at the centers, then the existence of
    solutions with two centers is not restricted to particular lines in the
    ($\eps, \Gamma$) plane. There will be a whole area of such solutions,
    bounded by $\eps= \min (\eps_*(\Gamma), -1)$ from below, and by the
    line $\eps=\eps_{h}(\Gamma)$ from above. Among the nonsymmetric
    solutions of this kind with several knots of the scalar field one can
    find those with multiple local maxima and minima of $\gamma(l)$.
    Their possible connection with matter trapping and the hierarchy problem
    is discussed below.

\section{Matter in the background of global string configurations}

    In this section we discuss the problem of trapping of classical point
    particles and test scalar fields by the gravitational field of the
    global string configurations described above.

\subsection{Classical particles}

    The motion of classical particles in the bulk may be equivalently
    described in terms of geodesics or using the Hamilton-Jacobi equation.
    We here use the second approach.

    The Hamilton-Jacobi equation for a point test particle of (primary) mass
    $m_0$ in space-time with the metric (\ref{Metric}) is
\[
    \e^{-2\gamma (l)} \biggl[ \left(\frac{\d S}{\d t}\right)^2
    -\left( \frac{\d S}{\d \vec{x}}\right)^2 \biggr]
    -\left( \frac{\d S}{\d l}\right)^2
    -\frac{1}{r^2 (l) }\left( \frac{\d S}{\d \theta }\right) ^2 -m_0^2 =0,
\]
    The metric is homogeneous with respect to all coordinates
    except $l$, and the action can be written in the form
\beq
    S = Et - \vec{p}\vec{x} + S_{l}(l) + M \theta, \label{Action}
\eeq
    where $E$ is the particle energy, $\vec{p}$ is the particle momentum
    along the coordinates $x^i$, $i = \overline{1, d_0-1}$,
    $\theta $ is the angular coordinate in the extra dimensions (note that
    $d\Omega^2 = d\theta^2$ for $d_1=1$) and $M$ is its conjugate angular
    momentum. The remaining unknown function  $S_{l}(l)$ satisfies the
    equation
\[
    \frac{dS_l}{dl}=\pm \sqrt{p^2 \e^{-2\gamma (l) }-\frac{M^2 }{r^2
                (l)}-m_0^2},
\]
    where $p^2 =E^2 -\vec{p}^2 $. Zeros of the square root determine the
    turning points of classical motion.

    Consider a particle with $M=0$, i.e., moving in the bulk along the
    coordinate $l$ (strictly to or from a brane if the latter is located at
    fixed $l$). Classical motion is allowed where the square root is real.
    The turning points $l_{t}$ are determined by the equation
\[
        p^2 \e^{-2\gamma (l_{t}) } - m_0^2 =0.
\]

    If there is a minimum of $\gamma(l)$ at some $l=l_0$, a classical
    particle with $p^2 = m_0^2  \e^{2\gamma (l_0) }$ cannot move along the
    $l$ direction and is trapped precisely at this minimum of $\gamma$.
    Particles with slightly larger $p^2$ can move between two turning points
    in the vicinity of $l_0$. If it is a global minimum of $\gamma$,
    particles with any $p^2 \geq m_0^2 \e^{2\gamma(l_0)}$  are trapped.

    It can also be verified that particles with the same value of $p^2$ but
    $M\ne 0$ (moving in the $\theta$ direction) have a still more narrow
    range of motion along $l$.

    In particular, near the equator of a configuration with two centers and
    three half-waves of $\phi$ (see Fig.\,\ref{2kn gamma(l)}), the turning
    points of finite classical motion exist for $\gamma (l_\q) <\gamma
    <\gamma_m $, where $\gamma_m $ is the maximum of $\gamma (l)$. Denoting
    $m_{\rm eq}^2 =m_0^2 \e^{2\gamma ( l_\q) }$, we see that a classical
    particle is trapped near the equator if its energy is restricted by
\[
       m_\q^2 < p^2 < m_\q^2 \e^{2[ \gamma_m -\gamma (l_\q)]}.
\]
    It moves along the Minkowski coordinates as a free particle of mass $m_\q$.

\subsection{Scalar fields \label{7.2}}

   Consider a test scalar field $\chi$ with the Lagrangian $L_\chi$ such that
\beq
      2L_\chi = \d^A \chi^* \d_A \chi - m_0^2 \chi^* \chi      \label{L-chi}
\eeq
   in the background of our string configurations with the metric
   (\ref{Metric}). Here, the asterisk as a superscript denotes complex
   conjugation and $m_0$ is the initial field mass. The $\chi$ field obeys
   the Klein-Gordon equation
\beq
    \d_A\left( \sqrt{g} g^{AB}\d_B\chi \right)
               + \sqrt{g} m_0^2 \chi =0,                \label{KG}
\eeq
   where $g = |\det (g_{AB})| =\e^{2d_0\gamma + 2\beta}$. Taking into account
   the symmetry of the problem, we can take a single mode of $\chi$, assuming
\beq \wide
       \chi (x^A) = X(l) \e^{-ip_\mu x^\mu}\,\e^{in\theta},     \label{mode}
\eeq
   where $p_\mu = (E, \vec p)$ is the $(d_0=4)$-momentum along the brane and
   $n$ is an integer. Then $X(l)$ obeys the equation
\beq
    X'' + (d_0\gamma' + \beta')X'                         \label{eq-X}
         + (p^2 \e^{-2\gamma} - n^2\e^{-2\beta} - m_\chi^2) X =0,
\eeq
   where $p^2 = p_\mu p^\mu = E^2 - \vec p{}^2$ is the effective mass
   squared, observed on the brane.

   As a trapping criterion for a mode $X$, it is reasonable to require
   finiteness of the $\chi$ field energy $E_\chi$ per unit area of the
   brane,
\beq
     E_\chi = \int T^t_t [\chi] \sqrt{g}\, d\theta\, dl        \label{E-chi}
            =  2\pi \int T^t_t [\chi] \sqrt{g}\, dl  < \infty,
\eeq
   where
\beq                                                         \label{T00}
     T^t_t [\chi] = \Half \left[\e^{-2\gamma} (E^2 + \vec p{}^2)
            + X'{}^2 + (n^2 \e^{-2\beta} + m_0^2)X^2 \right]
\eeq
   is the temporal component of the $\chi$ field stress-energy tensor.
   One can notice that the validity of (\ref{E-chi}) automatically
   guarantees finiteness of the norm $\int \sqrt{g} \chi^*\chi\, dl\,
   d\theta$ of the $\chi$ field considered as a quantum-mechanical wave
   function.

   The finiteness of $E_\chi$ in the background of different regular
   configurations with infinite extra dimensions described above depends on
   the behavior of solutions to (\ref{eq-X}) at small and large $l$.

   We begin with considering the $\chi$ field behavior near a regular
   center $l=0$, which is common to all classes of regular configurations.
   At small $l$, we have $\e^\beta \equiv  r \sim l$ and $\gamma\to 0$.
   Hence \eq (\ref{eq-X}) takes the approximate form
\bear
     l X'' + X' + l(p^2 - m_0^2) X \eql 0 \cm (n=0),    \label{n=0}
\\
     l X'' + X' - (n^2/l) X \eql 0 \cm  (n\ne 0).       \label{n>0}
\ear
   \eq (\ref{n=0}) is solved by zero-order cylindrical functions if $p^2\ne
   m_0^2$ and in elementary functions if $p^2 = m^2$; \eq (\ref{n>0}) is an
   Euler equation. The solutions behave as follows at small $l$:
\bear
        X \al \approx \al C_1 + C_2 \ln l \cm (n=0),    \label{l_to_0}
\nn
        X \al \approx \al C_3 l^n + C_4 l^{-n} \cm (n\ne 0),
\ear
   with integration constants $C_i$. To make the integral in (\ref{E-chi})
   converge as $l\to 0$, one should choose $C_2 =0$ and $C_4=0$, i.e., in
   each case, only one of the two linearly independent solutions.

   Now, consider the asymptotic form of solutions to \eq (\ref{eq-X}) as
   $l \to \infty$ for different background configurations.

\medskip\noi
   {\bf A1:} at large $l$, $\gamma \sim \beta \sim hl$, $h = \const  >0$. In
   \eq (\ref{eq-X}), the terms with $p^2$ and $m^2$ are negligible, and
   the solution has the asymptotic form
\beq
      X \approx C_+ \e^{-a_+ l} + C_- \e^{-a_-l}, \cm
     2a_\pm = (D-1)h \pm \sqrt{(D-1)^2 h^2 + 4m_0^2},     \label{XA1}
\eeq
   where $C_\pm$ are integration constants and $D=d_0+2$ is the full
   space-time dimension. It is easy to verify that the criterion
   (\ref{E-chi}) holds for the solution with $C_+ \ne 0$, $C_-=0$. Thus
   scalar fields with any nonzero mass can be trapped on such branes.

\medskip\noi
   {\bf A2:} at large $l$, $\e^{d_0\gamma} \sim \e^{\beta} \sim l$. Again,
   the terms with $p^2$ and $m^2$ are negligible, and \eq (\ref{eq-X})
   transforms to $X'' + 2X'/l -m_0^2 X =0$ whose solution is
\beq
     X \approx (C_+ \e^{m_0 l} + C_- \e^{-m_0 l})/l,           \label{XA2}
\eeq
   and evidently the solution with $C_+ =0$ satisfies the criterion
   (\ref{E-chi}).

\medskip\noi
   {\bf B1:} at large $l$, $r\equiv \e^\beta \to r_*$, $\gamma \approx -hl$,
   $h>0$, and the approximate form of \eq (\ref{eq-X}) is
\beq
    X'' - d_0 h X' + p^2 \e^{2hl} X =0.
\eeq
   For $p \ne 0$, it is solved in cylindrical functions, the general
   solution being
\beq                                                          \label{XB1}
    X = \e^{d_0 hl/2}Z_{d_0/2} \biggl(\frac{|p|}{h} \e^{hl}\biggr)
        \sim \e^{(d_0-1)hl/2} \sin \biggl(\frac{|p|}{h} \e^{hl} +\Phi\biggr),
\eeq
   where $\Phi$ is a constant phase. It is easy to verify that $E_{\chi}$
   diverges as $\int \e^{hl}\, dl$. So massive modes with any $p^2 >0$
   are not trapped by B1 configurations.

\medskip\noi
   {\bf B2:} at large $l$, $r \to r_*$ and $\gamma \approx hl$, $h > 0$.
   The situation is almost the same as in case A1; the solution to
   (\ref{eq-X}) has the asymptotic form (\ref{XA1}) with the replacements
\[
     D-1 \mapsto d_0, \cm m_0^2 \mapsto m_0^2 + n^2/r_*^2.
\]
   Again, only the solution with $C_- =0$ provides convergence of $E_\chi$.

\medskip
   Thus the configurations of classes A1, A2 and B2 can trap massive scalar
   modes; since at both large and small $l$ only one of the two linearly
   independent solutions to (\ref{eq-X}) is selected, we have a
   boundary-value problem with a discrete spectrum of $p^2$ for any given
   values of $m_0$, $n$ and the background parameters.

\subsection{The Schr\"odinger equation}

   It is helpful to reformulate the boundary-value problem for scalar field
   modes in terms of the Schr\"odinger equation. To do so, we make the
   following substitutions in (\ref{eq-X}):
\beq
     dl = \e^{\gamma} dx, \cm X(l) = y(x)/\sqrt{f(x)},      \label{X_to_y}
\eeq
   where $f(x) = \e^{(d_0-1) \gamma + \beta}$. The new variable $x$ is
   actually an analogue of the well-known ``tortoise coordinate'' in the
   analysis of \sph\ metrics, such that the metric takes the form
\beq
    ds^2 = \e^{2\gamma}\big (\eta\mn dx^\mu dx^\nu - dx^2 \big )
            - \e^{2\beta}\, d\theta^2.          \label{ds-x}
\eeq
   Then \eq (\ref{eq-X}) transforms to the Schr\"odinger form
\beq
    y_{xx} + [p^2 - V\eff (x)] y =0                    \label{Schr}
\eeq
   with the effective potential
\beq                                                        \label{V_eff}
    V\eff =  \big( m_0^2 + n^2 \e^{-2\beta} \big) \e^{2\gamma}
        + \frac{f_{xx}}{2f} - \frac{f_x^2}{4f^2},
\eeq
   where the subscript $x$ denotes $d/dx$. Recall that the eigenvalue $p^2$
   is the effective mass squared, observed in Minkowski space.

   Near the center (without loss of generality, $x \approx l \to 0$), we have
\beq
    V\eff \approx n^2/x^2 + m_0^2 + \frac{1}{4}
        \biggl[1 + 2(d_0-1)\gamma_{xx} + \beta_{xx}\biggr]_{x=0}.
\eeq
   So for $n\ne 0$ it is a potential wall whereas $V\eff \to \const$ for
   $n=0$.

   At large $l$ for different backgrounds, we have:
\bear   \nq
   {\bf A1:} \quad && x\to x_+ < \infty; \quad x_+ -x\sim \e^{-hl}, \quad h>0;
\nn
       &&  V\eff(x) \approx \e^{2hl}                        \label{VA1}
            \biggl[ m_0^2 + \frac{h^2}{4}(d_0^2 + 2d_0) \biggr].
\yy
   {\bf A2:} \quad && x\sim l^{(d_0-1)/d_0}\to \infty;
\nn                                                           \label{VA2}
       &&  V\eff(x) \approx m_0^2 \e^{2\gamma}
                    \sim l^{2/d_0}\sim x^{2/(d_0-1)}.
\yy
   {\bf B1:} \quad && x \sim \e^{hl }\to \infty, \quad h>0;    \label{VB1}
\nn
       &&  V\eff(x) \approx \e^{-2hl}
            \biggl[ m_0^2 + \frac{n^2}{r_*^2} +
            \frac{h^2}{4}(d_0^2 -1) \biggr] \sim \frac{1}{x^2}.
\yy
   {\bf B2:} \quad && x\to x_* < \infty; \quad x_*-x \sim \e^{-hl},\quad h>0;
\nn
       &&  V\eff(x) \approx \e^{2hl}                          \label{VB2}
              \biggl[ m_0^2 + \frac{n^2}{r_*^2} +
                    \frac{h^2}{4}(d_0^2 -1) \biggr].
\ear


   We see that in the cases A1, A2, B2 the potential is rising to infinity
   at large $l$, which leads to discrete spectra of $p^2$. For B1
   configurations (with a horizon at $l=\infty$), in usual quantum
   mechanics we would expect a continuous spectrum of states; in our case,
   with the appropriate boundary conditions, as we saw above, there are no
   admissible states with $p^2 > 0$.

\subsection {Massless modes in configurations with infinite extra dimensions}

   For a possible massless mode, $p^2 = m_0^2 = n^2 =0$, \eq (\ref{eq-X}) is
   easily solved:
\beq                                                         \label{0-mode}
      X' = C_1 \e^{-d_0 \gamma - \beta}, \cm
      X = C_1 \int \e^{-d_0 \gamma - \beta} dl + C_2, \cm C_{1,2} = \const.
\eeq
   and $X$ is found by quadrature.

   One of the solutions is $X=\const$. It is easy to verify that with this
   solution, which is well-behaved at a regular center, the energy $E_\chi$
   (\ref{E-chi}) diverges at large $l$ in the backgrounds A1, A2, B2 but
   converges in the background B1.

   For the other solution with $C_1 \ne 0$, on the contrary, $E_\chi$
   converges at large $l$ in the backgrounds A1, A2 and B2 and diverges in
   B1. This solution, however, is singular at the center and leads there to
   a divergence in $E_\chi$.

   Thus B1 configurations with horizons, being unable to trap massive scalar
   fields, are the only ones that can trap a massless scalar.

\section{Configurations with two centers and the hierarchy problem}

   In configurations with two symmetric regular centers and two knots of the
   scalar field, there are three minima of the ``gravitational potential''
   $\gamma$, see Fig.\,\ref{2kn gamma(l)}. The minimum at the equator is
   higher than the other two, located near the centers. A similar (and even
   more complicated) structure may be expected for configurations with a
   larger number of scalar field knots. The minima of $\gamma$ are able to
   trap classical particles. As to quantum ones (at least for zero spin),
   the effective potential (\ref{V_eff}) not necessarily has a minimum
   precisely where $\gamma$ has a minimum, and an additional detailed study
   is necessary. Though, at least quasiclassically, quantum and classical
   particles must be trapped in close positions, and the main difference
   between them is that quantum particles can tunnel from a higher minimum
   of $V\eff$ to a lower one.

   Now, suppose a particle described by a certain mode of the $\chi$
   field (for simplicity, with $n=0$) is trapped at some position $l_i$. The
   mode equation (\ref{eq-X}) may be rewritten in the form
\beq                                                        \label{eq-X1}
      \left( \sqrt{g} X' \right)' + \sqrt{g} X \e^{-2\gamma} p^2 =
            \sqrt{g} m_0^2 X.
\eeq
   Let us integrate this equation over the extra dimension from one center
   to the other. The term $\int \left( \sqrt{g} X' \right)' dl=0$ because
   $\sqrt{g} = r\e^{d_0 \gamma }$ is zero at both centers. For a particle
   trapped at some fixed position $l=l_i$, we get
\beq                                                        \label{m_i}
    p^2 = m_i^2 = m_0^2
    \frac{\int \sqrt{g} X \,dl}{\int \sqrt{g}\e^{-2\gamma }X\,dl}
            \approx m_0^2 \e^{2\gamma (l_{i})}.
\eeq

   To interpret this result, we note that the whole picture looks quite
   different depending on the size of the extra dimensions, characterized by
   the distance $l_c$ between the centers. This size, in turn, varies with
   the value of $\Gamma = \kappa^2 \eta^2$: it is close to unity (i.e., the
   length unit which is also arbitrary) for large $\Gamma$ and tends to
   infinity as $\Gamma \to 0$. For (comparatively) weak gravity of the
   string, $\Gamma \to 0$, when $l_c$ is very large, all minima of $\gamma
   (l) $ form individual branes located in the bulk far from one another. In
   this case, an observer located on one of the branes sees only particles
   corresponding to modes trapped on this brane; tunnelling from one brane
   to another will be seen as appearance or disappearance of observable
   particles. The whole picture may be used for treating the interaction
   hierarchy problem in the spirit of Randall and Sundrum's first model
   \cite{RS1}.

   In the opposite case $\Gamma \to \infty $ (and provided the unit
   length $(\lambda_0 \eta)^{-1/2}$ is also sufficiently small), $l_c$ can
   be a length invisible for modern instruments, say, $l_c \ll 10^{-17}$ cm.
   We then arrive at a picture close to the original Kaluza-Klein concept;
   particles with the same primary mass $m_0$, being trapped at different
   minima of the effective potential, are seen as particles with different
   masses, and the tunneling process from a higher minimum to a lower one is
   observed as a decay of a particle of larger mass to that of smaller mass,
   with energy release in some form. This may be a natural explanation of
   the existing families of particles with different masses but similar
   other properties. A more detailed study of this opportunity is desirable
   but is beyond the scope of the present paper.

\section {Conclusion}

    Our phenomenological approach based on the macroscopic theory of
    phase transitions with spontaneous symmetry breaking allows us to
    study the general physical properties of topological defects in the
    framework of the brane world concept. In particular, in this paper we
    have studied the gravitational properties of global strings located in
    extra dimensions. We have given a general description and classification
    of possible regular solutions and presented a map showing the location
    of different solutions in the space of physical parameters.

    Among the variety of regular solutions, there are ones possessing
    brane features, including those with multiple branes, as well as those
    of potential interest from the viewpoint of the hierarchy problem.

    In connection with branes, we have analyzed the possibilities of
    trapping of classical particles and scalar fields. We have shown that,
    contrary to a domain wall, in the case of an extra-dimensional global
    string, matter can be trapped by gravity even without coupling to the
    scalar field that forms the string itself.

    Among the configurations with two centers, the structures having several
    minima of $\gamma(l)$ may be of interest in connection with the
    hierarchy problem. If the distance between the centers is small, we work
    within the Kaluza-Klein concept, and the same particle, being trapped at
    different minima, looks for an observer as a family of similar
    particles with different rest masses.

\Acknow{We appreciate partial financial support from Russian Basic
    Research Foundation Project No. 05-02-17478. }

\end{document}